\documentclass[%
 aip,
 amsmath,amssymb,
 reprint,%
]{revtex4-1}

\usepackage{graphicx}
\usepackage{dcolumn}
\usepackage{bm}

\usepackage[utf8]{inputenc}
\usepackage[T1]{fontenc}
\usepackage{mathptmx}
\usepackage{etoolbox}
\usepackage{xcolor}
\usepackage{braket}
\usepackage{dsfont}
\DeclareMathOperator{\tr}{Tr}
\DeclareMathOperator{\id}{{\mathds 1}}
\usepackage{calrsfs}
\DeclareMathAlphabet{\pazocal}{OMS}{zplm}{m}{n}

\usepackage{soul}

\makeatletter
\def\@email#1#2{%
 \endgroup
 \patchcmd{\titleblock@produce}
  {\frontmatter@RRAPformat}
  {\frontmatter@RRAPformat{\produce@RRAP{*#1\href{mailto:#2}{#2}}}\frontmatter@RRAPformat}
  {}{}
}%
\makeatother
\begin{document}

\preprint{AIP/123-QED}

\title{Quantum thermodynamics under continuous monitoring: \\ a general framework}
\author{Gonzalo Manzano}

\author{Roberta Zambrini}%
 \email{gonzalo.manzano@ifisc.uib-csic.es}
\affiliation{Institute for Cross Disciplinary Physics and Complex Systems (IFISC) UIB-CSIC, Campus Universitat Illes Balears, Palma de Mallorca, Spain.
}%


\begin{abstract}
The thermodynamics of quantum systems driven out of equilibrium has attracted increasing attention in last the decade, in connection with quantum information and statistical physics, and with a focus on  non-classical signatures. While a first approach can deal with average thermodynamics quantities over ensembles, in order to establish the impact of quantum and environmental fluctuations during the evolution, a continuous quantum measurement of the open system is required.
Here we provide an introduction {to the general theoretical framework} to establish and 
interpret thermodynamics for quantum systems whose nonequilibrium evolution is continuously monitored. We review the formalism of quantum trajectories and its consistent application to the thermodynamic scenario, where main quantities such as work, heat, and entropy production can be defined at the stochastic level. The connection to irreversibility and fluctuation theorems is also discussed, {together with some recent developments,} and we provide some simple examples to illustrate the general theoretical framework.
\end{abstract}

\maketitle

\section{\label{sec:intro} Introduction}

Quantum thermodynamics is a growing and rapidly-evolving field at the intersection of quantum information, many-body physics and nonequilibrium thermodynamics that has attracted a great deal of attention in the last decade~\cite{Goold2016,Vinjanampathy2016}. It aims to describe work, heat and entropy production along quantum nonequilibrium processes with a special attention to genuine quantum phenomena. Paradigmatic examples include studying the thermodynamic role of quantum coherence~\cite{Skrzypczyk2014,Lostaglio2015,Kammerlander2016,Santos2019,Francica2019} also in view of applications~\cite{Scully2010,Poem2019,Manzano2020,Latune2021}, quantum correlations like entanglement~\cite{Oppenheim2002,Hovhannisyan2013,Perarnau2015} or discord~\cite{Zurek2003,Francica2017,Manzano2018}, addressing the effects of quantum measurements~\cite{Jacobs2012,Elouard2017,Buffoni2019,Guryanova2020,Debarba2019}, and exploring the link between energy and (quantum) information~\cite{Sagawa2009,Rio2011,Horodecki2013,Bera2019,Lostaglio2019}. While most works in the field until now have focused on first-principle definitions and the behavior of average thermodynamic quantities, fluctuations are gaining increasing attention in recent years. Classical and quantum fluctuations are indeed known to be at the core of thermodynamic behavior at small scales, where genuine trade-offs and universal nonequilibrium relations constraining energetic and entropic quantities emerge~\cite{Jarzynski2011,Seifert2012,Horowitz2019}. In this context, the framework of quantum trajectories and related methods describing the indirect and continuous monitoring of quantum systems, provides an ideal platform to explore stochastic thermodynamics in the quantum regime. 

Quantum trajectories were first considered in quantum optics~\cite{Dalibard1992,Carmichael1993} to describe processes such as photodetection, and to simulate the dynamics of open quantum systems when the master equation approach becomes intractable~\cite{Plenio1998,Daley2014}. Nowadays quantum trajectories are generated and recorded in the laboratory in number of different platforms ranging from superconducting few-level systems~\cite{Vijay2011,Murch2013,Vool2014,Huard2016,Minev2019} to optomechanical setups~\cite{Aspelmayer2015,Rossi2019}, including pioneering experiments with trapped ions~\cite{Wineland1986} and cavity QED platforms~\cite{Gleyzes2007,Haroche2013}. Its use have been proposed for different scopes including quantum state estimation~\cite{Molmer2013,Guevara2015} and control~\cite{Wiseman2009,Sayrin2011,Roch2014}, detection of dynamical phase transitions~\cite{Garrahan2010} and the characterization of quantum synchronization~\cite{Najmeh2020,Weiss2016} among others. The framework for the characterization of stochastic and quantum thermodynamics along quantum trajectories that we  {introduce} here  {has} been {roughly} developed during the last decade,  {and is capturing}  increasing attention. Its starting point can be situated in the pioneering efforts of J. M. Horowitz~\cite{Horowitz2012} and of F. W. J. Hekking and J.P. Pekola~\cite{Hekking2013} to study and interpret the quantum jump approach in thermodynamic terms for particular representative cases (a driven dissipative harmonic oscillator and a driven two-level system). These {two} works were based, at the same time, in previous studies that obtained partial but useful results, see e.g. Refs.~\cite{Breuer2003,Roeck2006,Derezinski2008,Crooks2008}. 
{C}ontributions from several groups within the community working in quantum thermodynamics~\cite{Horowitz2013,Leggio2013,Liu2014,Horowitz2014,Suomela2015,Manzano2015,Liu2016,Auffeves2017,Elouard2017b,Manzano2018b,Gherardini2018b,Mohammady2020,Miller2021,Carollo2021} 
generalized and tested the framework {in the last 8 years}, including extensions to scenarios with feedback control~\cite{Strasberg2013,Gong2016,Murashita2017,Naghiloo2018,Strasberg2019}, diffusive noise~\cite{Alonso2016,Auffeves2017,DiStefano2018,Belenchia2020,Rossi2020} and arbitrary environments~\cite{Manzano2018b}. Applications to quantum heat engines~\cite{Campisi2015,Liu2020,Brandner2020}, probing correlations~\cite{Gherardini2018,Elouard2019} and the erasure of information~\cite{Miller2020,Manzano2018c} have been proposed, as well as the development of experimental proposals for measuring heat and work along individual  trajectories~\cite{Pekola2013,Suomela2016,Donvil2018,Naghiloo2020,Karimi2020}. Recently, the framework has been also used to obtain generalized versions of the Thermodynamic Uncertainty Relations (TUR)~\cite{Carollo2019,Hasegawa2020,Hasegawa2021,Miller2021b} ---that establish trade-off relations between the fluctuations of observable currents and dissipation--- and to develop a Quantum Martingale Theory (QMT) describing the thermodynamics of processes at stopping times (such as first-passage times or escape times) in connection to quantum features~\cite{Manzano2019,Manzano2021}.


Within the quantum trajectory approach there exist different ways to handle the continuous measurement schemes: the so-called unravellings, the possibility of {efficient or inefficient detection}, etc. This leads to a variable difficulty to identify the relevant thermodynamic quantities {at} 
different levels of generality. Moreover there has been often different proposals for the interpretation of the thermodynamic quantities arising in the framework and their interplay. In particular, the identification of energetic fluctuations due to measurement backaction as either work or heat have raised an ongoing debate in the community, as we will address in more details later. Nevertheless, this collaborative effort has provided a powerful {and promising} extension of stochastic thermodynamics to the quantum realm. This extension not only allows to apply the general understanding and inference possibilities of stochastic thermodynamics to small systems where quantum features cannot be neglected, but it may also help to unveil genuine thermodynamic features of quantum coherence and correlations, and provides new insights to our fundamental understanding of quantum measurements.

In the present review {we focus on the theoretical framework providing} an accessible 
overview of the main ingredients needed to establish and interpret thermodynamics {of} quantum systems whose nonequilibrium evolution is continuously monitored. {We propose a route starting from central concepts extended from classical stochastic thermodynamics and quantum thermodynamics of isolated systems, which are extended to the quantum trajectory scenario.   Here the concept of microreversibility in the evolution will play a central role, over which the whole framework is constructed.} {In order to provide a balanced presentation in some sections we extend the formalism to situations not systematically treated in the literature, as for diffusive trajectories where microreversibility issues may arise (Sect. \ref{subsec:micro-diffusive} C}).
We then discuss the energetics of quantum trajectories  {in Sect.\ref{sec:energetics}, reviewing} different proposals made in the literature and clarify some 
points needed to reach a solid understanding and coherent interpretation of the {main thermodynamic} quantities. {The review is again complemented with an extension of} the framework {in oder to  accommodate} situations with multiple conserved quantities and discuss in details some important points such as the assessment of irreversibility through entropy production and their fluctuations, as well as different possible splits into contributions that provide extra insights on the thermodynamic behavior of the system and their genuine quantum properties {(Sects.\ref{sec:ep} and \ref{dec_ep})}. Some aspects of the general framework are illustrated in two simple examples, {Sect. \ref{sec:examples},} while some first experiments that started to explore thermodynamics of quantum trajectories {are mentioned}, together with other promising platforms. Finally, we provide an outlook on further possible developments and their applications in the field.

\section{\label{sec:quantumtrajectories} Quantum trajectories in a nutshell}

Quantum trajectories describe the evolution of systems monitored through selective measurement and provide a powerful approach to the description of open quantum systems. Both methodological convenience and the need of a theoretical description accounting for continuous monitoring have been driving motivations for developing this framework. Indeed, with quantum trajectories, numerical simulation of master equations requires less memory and time, relying on  (stochastic evolution of) pure states instead of density matrices and being naturally adapted to parallel computation. More crucially, quantum trajectories fill the theoretical gap between  the unitary evolution  of isolated systems and the master equation evolution in presence of large and oblivious environments, accounting instead for the distinctive effects of selective measurement. This topic is presented in detail in Refs. \cite{GardinerZoller,Wiseman2009,Percival1998,Barchiellibook,Brun2002,Jacobs2006}, 
to name a few. In the following we briefly introduce the framework for both  quantum-jump and diffusive trajectories, that correspond to the most common sets of experimental records, namely discontinuous (point) events or continuous signals.

We consider a system monitored though a continuous generalized measurement, 
changing the state of the system {at each small time step $dt$}.  Positive operator value measurements are defined considering a set of $(K+1)$ operators $\Omega_k$ such that $\sum_k \Omega^\dagger_k\Omega_k=\id$ \footnote{This can be also generalized to a continuous set of operators \cite{Jacobs2006}}. Depending on the experimental setting, the operators $\Omega_k$ can be {sharp projections or {other measurements like POVM operators, including} smooth projectors superpositions}~\cite{Wiseman2009}. For a given outcome {$k$} of {the} (selective) measurement {at time $t$}, the corresponding {updated} state of the system {becomes}  ${\rho_t \rightarrow~} \Omega_k \rho_t \Omega^\dagger_k/P_k$ with $P_k=\tr[\Omega_k \rho_t \Omega^\dagger_k]$ {the probability to obtain outcome $k$ in the measurement}. The state after unselective measurement, that is considering the ensemble mixture of measurement outputs, is therefore $\rho_{t + dt} = \sum_k \Omega_k \rho_t \Omega^\dagger_k$. 

{In the quantum trajectories framework, the main focus is on describing the continuous monitoring of open quantum systems following Markovian evolution. Therefore, i}f the measurement outcome is not selectively monitored, the change rate of the state  $[\rho_{t+dt}-\rho_{t}]/dt$ 
induced by this $unselective$ measurement in the limit $dt\rightarrow0$ {is assumed to} correspond to a Lindblad [or Gorini-Kossakowski-Sudarshan-Lindblad (GKLS)] master equation \cite{Wiseman2001}
\begin{equation} \label{eq:Lindblad}
\dot{\rho}=\mathcal{L}(\rho) = -i[{H}, \rho] + \sum_{k=1}^K {L}_k \rho {L}_k^\dagger - \frac{1}{2}\{{L}_k^\dagger{L}_k, \rho \}
\end{equation}
{where $H$ is an Hermitian operator corresponding to the monitored system Hamiltonian, and $L_k$ are the so-called Lindblad operators.}
The non-unitary part of the dynamics is modeled by the sum term in Eq.~(\ref{eq:Lindblad}), with dissipators $\mathds{D}_k(\rho) \equiv L_k \rho L_k^\dagger - \frac{1}{2}\{L_k^\dagger L_k, \rho\}$. {Although not explicitly written in the above equation, we will also generically allow for temporal dependencies in the operators appearing in the master equation~\eqref{eq:Lindblad}. In particular, we consider that $H(\lambda)$ {and $L_k(\lambda)$} might depend on a control parameter $\lambda(t)$ that can vary in time, which allow us to model driving processes following externally operated protocols. Such protocols will be important in the following sections, and in that case we will write the Lindbladian in Eq.~\eqref{eq:Lindblad} as $\mathcal{L}_\lambda (\rho)$.}
In the next, we will instead define the evolution corresponding to a given measurement record, known as quantum trajectory.

\subsection{Quantum-jump trajectories}

{A first important class of quantum trajectories in which we focus is known as ``quantum-jump'' trajectories, which are obtained} for a set of measurement operators (complete up to first order in $dt $)
\begin{eqnarray}\label{eq:measOp}
\Omega_0&=&\id - \frac{dt}{2}\sum_{k\neq0}  L^\dagger_k L_k - idtH, \\
\Omega_k&=&\sqrt{dt} L_k ~~~~\text{ for } 1\le k \le K, \nonumber
\end{eqnarray}
with $dt\ll 1$. 
{Here} the state {of the system} will be only weakly modified for measurement record $k=0$, an event occurring with probability $P_0 \sim1$. On the other hand, for the other outcomes $k\neq0$, a substantial change will occur in the system state, but the probability  will be negligible ($P_k\sim dt$). This is the largely explored quantum optics scenario in which a jump is detected, corresponding to a photon emission from a decaying atom {modeled by $L_k= \sigma_{-}$}. It corresponds to a measurement record taking either values 0 (more frequently) or 1 (when a detector clicks). 

The sequence of records of such a measurement in time is denoted by $\gamma_{(0,\tau)} {= \{dN(t) ; 0\leq t \leq \tau\}}$, and constitutes a realization of a stochastic process, where $dN(t)=\{0,1\}$ is a stochastic increment corresponding to either no-click or one click in the detector during the interval $[t, t+ dt]$. The number of clicks in the detector up to time $\tau$ is hence $N(\tau) = \int_{0}^\tau dN(t)$. In the more general case considered here, with $K$ distinguishable {channels} (like e.g. energy lowering and raising processes or emission of photons in different modes) the measurement record $\gamma_{(0,\tau)}$ includes $K$ stochastic increments $dN_k(t)={\{0 ,1\}}$, each of which ``{signaling}'' when a jump of type $L_k$ is detected. Being the probability of a joint event negligible, we assume $dN_n(t) dN_k(t)=\delta_{n k} dN_k(t)$. The (classical) averages of the record of measurements $\langle\cdot \rangle_\gamma$ can be associated to the quantum expectation values   $\langle\cdot \rangle = \tr[\rho \cdot]$ of the corresponding measurement operators, so that {the probability of a jump in the interval $[t, t + dt]$ reads}
\begin{equation}  
\langle dN_k(t)\rangle_\gamma=\langle \Omega^\dagger_k\Omega_k \rangle=dt\langle L^\dagger_kL_k \rangle  ~~~~~~~~ k\neq 0,
\end{equation}
{which} tell us the rate at which jumps of type $k$ occur during the evolution. {Since this probability only depends on quantities evaluated at time $t$, the statistics of the jumps are Poissonian with (time-dependent) intensity $\langle L^\dagger_kL_k \rangle$.}

If we now consider the evolution of the {pure} state of a system {$\ket{\psi(t)}$} continuously monitored through the measurement \eqref{eq:measOp}, depending of the detection $dN_k(t)$, the updated state will correspond to $\Omega_0 \ket{\psi}$ or $\Omega_{k} \ket{\psi}$ with $k\neq0$, respectively. {The state increment $d|\psi(t)\rangle=|\psi(t+dt)\rangle-|\psi(t)\rangle$ can then be constructed by combining these possibilities, each of which multiplied by the factor $[1 - \sum_{k\neq 0} N_k]$ (which becomes 0 when a jump is detected and is 1 otherwise) and the increments $dN_k$, respectively. Taking the limit $dt \rightarrow 0$}
one obtains the nonlinear stochastic Schr\"odinger equation {(SSE)} for the conditional dynamics of the monitored state \cite{Wiseman2009}
\begin{eqnarray} \label{eq:SSE}
d |\psi_\gamma(t)\rangle&=&-iH dt |\psi_\gamma(t)\rangle - 
 \frac{dt}{2} \sum_k \left(L^\dagger_k L_k - \langle L^\dagger_k L_k\rangle  \right) |\psi_\gamma(t)\rangle \nonumber\\
 &+& \sum_k dN_k(t) \left(  \frac{L_k}{\sqrt{\langle L^\dagger_k L_k\rangle}}    - \id  \right) |\psi_\gamma(t)\rangle
\end{eqnarray}
where terms $dN_kdt=o(dt)$ {have been} 
neglected {and $k=1, ..., K$ in the sums}. {As can be appreciated in the above equation, the evolution of the system is smooth and given by the first line when $dN_k(t) = 0 ~ \forall k$, while abrupt jumps in the system state occur whenever $dN_k = 1$ for some $k$, as given by the second line.}

The corresponding {stochastic} master equation {(SME)} for {a mixed conditioned} state $\rho_\gamma(t)$ {acquires a more compact form} (omitting time dependence)
\begin{eqnarray}\label{master_jumps}
    d \rho_\gamma  = -i  [H,\rho_\gamma] dt-\sum_k \left[\mathds{M}_k(\rho_\gamma)  {dt}
   -   \mathds{J}_k(\rho_\gamma) dN_k \right],
\end{eqnarray}
with drift (no-jump detection) terms 
\begin{eqnarray}
\mathds{M}_k(\rho_\gamma) =  {\frac{1}{2}}\{L^\dagger_k L_k, \rho_\gamma\} - \tr(L_k^\dagger L_k \rho_\gamma) \rho_\gamma
\end{eqnarray}
and jumps {super-operators}
\begin{eqnarray}
\mathds{J}_k(\rho_\gamma)=\frac{L_k \rho_\gamma L_k^\dagger}{\tr(L_k^\dagger L_k \rho_\gamma)} - \rho_\gamma. 
\end{eqnarray}
{We notice the SME above can be directly obtained from Eq.~\eqref{eq:SSE} by identifying $\rho_\gamma(t)=|\psi(t)\rangle\langle\psi(t)|$, however Eq.~\eqref{master_jumps} remains also valid for arbitrary mixed initial states of the system~\cite{Wiseman2009}}.

{Since the conditional system evolution consists of a sequence of smooth drift steps intersected by a set of $K$ rare jump events occurring with small probabilities $dt \langle L_k^\dagger L_k \rangle$, the measurement record $\gamma(0, \tau)$ can be alternatively given by specifying the times $t_j$ at which jumps of each time $k_j$ were detected:}
{
\begin{eqnarray}
{\gamma}_{(0,\tau)} = \{ (k_1,t_1), (k_2,t_2), ..., (k_J, t_J)\}, 
 \end{eqnarray}
{where we assumed a total number of $J$ jumps detected up to time $\tau$ {accounting for all $K$ channels},} while drift dynamics occurs the rest of the time. The {associated} evolution trajectory operator for the monitored state is
 \begin{eqnarray} \label{eq:T-operator}
 \pazocal{T}({\gamma}_{(0,\tau)}) = \mathcal{U}(\tau, t_J) L_{k_J} ... \mathcal{U}(t_2, t_1) L_{k_1} \mathcal{U}(t_1, 0),
\end{eqnarray}
with $\mathcal{U}(t_2, t_1)= \text{T}_+ \exp[-i \int_{t_1}^{t_2}dt(H  + i
 \sum L^\dagger_k L_k)]$ drift (or no-jump) operators modeling a smooth dynamics between times $t_1$ and $t_2$ and $\text{T}_+\exp$ time-ordered exponential (allowing for time dependent Hamiltonians or jumps $L_k$).
Notice that we are not enforcing state normalization here so that the probability of a given 
{measurement record over a initial state $\rho_0$ is given by $p_\gamma = \tr[\pazocal{T}^\dagger(\gamma) \pazocal{T}(\gamma)\rho_0]$}. The physical state at final time $\tau$ given a certain measurements record is then $\rho_\gamma(\tau)=\pazocal{T}(\gamma)\rho_0 \pazocal{T}^\dagger(\gamma) / p_\gamma$.}
{Finally, when averaging over measurement records $\gamma_{(0,\tau)}$, we recover the unconditional evolution of the system as determined by Eq.~\eqref{eq:Lindblad}, that is $\rho_\tau = \sum_\gamma \pazocal{T}(\gamma)\rho_0 \pazocal{T}^\dagger(\gamma) = \mathcal{E}(\rho_0)$, with $\mathcal{E} \equiv \text{T}_+ \exp[-\int_0^\tau dt \mathcal{L}]$.}

\subsection{Diffusive trajectories}

Let's consider now the case in which, monitored quantities produce one or more (K) {continuously} fluctuating signals, instead of discontinuous jumps, as it would occur with photocurrents or electrical currents, voltages, etc... In this case, the measurement records are continuous but not differentiable processes in time 
${\gamma_{(0,\tau)} }= \{I_k(t) ~;~ 0\leq t \leq \tau \}_{k=1}^K$, with $K$ {current} records at each time.
A diffusive {stochastic} evolution equation can be obtained in this case {reading} \cite{Wiseman2009,GardinerZoller,Percival1998}
\begin{align}\label{diff_eq_psi}
d |\psi_\gamma(t)\rangle &= \Big[-iHdt   - 
 \frac{dt}{2}\sum_k \left(L^\dagger_k L_k - 2\langle L^\dagger_k\rangle L_k  - {|\langle L_k\rangle|^2}\right) 
 \nonumber\\
 &~~~~~~+ \sum_k dw_k(t) \left({L_k}-\langle L_k\rangle\right) \Big] |\psi_\gamma(t)\rangle
\end{align}
 with $dw_k(t)$ a set of {real} valued Wiener increments with zero mean $\langle d w_k(t) \rangle = 0$ and non-vanishing correlations $\langle dw_k(t) dw_l(t^\prime) \rangle = \delta_{k l} \delta(t-t^\prime) dt$.
{Other similar versions of the diffusive SSE have been derived by many authors~\cite{Carmichael1993,Brun2002,Jacobs2006}, which may also include complex-valued Wiener increments~\cite{Wiseman2009}.}

{This equation can be derived from the previous quantum jump description {by using the symmetries of the Lindblad master equation. In particular Eq.~\eqref{eq:Lindblad} remains unchanged by a transformation in the Lindblad operators $L_k \rightarrow L_k^\prime = L_k + \alpha_k$, accompanied by a change in the Hermitian operator $H \rightarrow H^\prime = H - i \sum_k(L_k \alpha_k^\ast + L_k^\dagger\alpha)/2$. By considering a scheme where the jumps $L_k^\prime$ are detected and the parameters $\alpha_k$ are taken real and big enough, the dynamics of the system can be coarse-grained} over a time interval in which several individual jumps occur, but the evolution remains still smooth~\cite{Wiseman1993,Wiseman2009,Najmeh2020}. This occurs for a coarse grained time ${\Delta t} \sim \alpha_k^{-3/2}$ that leads to a number of detected jumps ${\Delta N_k} \sim \sqrt{\alpha_k}$ and to large number of counts {$(\sim \alpha_k^2)$}, consistent with a Gaussian statistics. {This is a prototypical case} in the quantum optics framework {when} moving to the limit in which the system optical mode {is} homodyne detected (superposed to a large coherent field $\alpha$, taken real). Similar situations arise as well in heterodyne detection~\cite{Wiseman2009}, while different classes of shifted operators like the $L_k^\prime$ above can be also obtained using chiral wavewides~\cite{Cilluffo2019}.
The measured record $\gamma_{(0,\tau)} = \{I_k(t) ; 0 \leq t \leq \tau \}$ in this homodyne{-like} detection scheme {becomes for} a variation during a small time interval $dt$,
\begin{eqnarray}
I_k(t) = \langle L_k + L_k^\dagger \rangle + dw_k(t)/dt,
\end{eqnarray}
{where $dw_k(t)/dt$ corresponds to {Gaussian} white noise}.
}

{Alternative approaches to diffusive processes start from a weak measurement framework modeled by a broad (unsharp) superposition of projectors [instead of Eq.~\eqref{eq:measOp}]. This allows for a more direct definition of a diffusion process and {some instances} can be found in Refs.~\cite{Jacobs2006,Brun2002}.}

{From the diffusive {SSE}~(\ref{diff_eq_psi}) a {diffusive SME} can be obtained {as well, reading}
\begin{eqnarray}
    d \rho_\gamma  =& -i&  [H,\rho_\gamma] dt + \sum_k \mathds{D}_k(\rho_\gamma) \\
    &+& \sum_k \left[(L_k -\langle L_k\rangle )\rho_\gamma+\rho_\gamma (L_k^\dagger -\langle L_k^\dagger\rangle )\right] dw_k(t), \nonumber
\end{eqnarray}
{with dissipators $\mathds{D}_k(\rho)$ as defined below Eq.~\eqref{eq:Lindblad}}, which is of the general form obtained in Refs.~\cite{Wiseman2001,Wiseman2009}. Interesting, both this and Eq.~(\ref{master_jumps})  are unraveling of the same GKLS master equation (\ref{eq:Lindblad}), when disregarding the measurement output, {as they correspond} to the same ensemble description.}

{Finally, the trajectory operators in this case can be written as a concatenation of measurement operators occurring at each infinitesimal time-step:
\begin{equation}
    \pazocal{T}(\gamma_{(0,\tau)}) = \Omega_{\mathbf{I}(\tau)} \Omega_{\mathbf{I}(\tau- dt)} \dots \Omega_{\mathbf{I}(t)} \dots  ~\Omega_{\mathbf{I}(dt)} \Omega_{\mathbf{I}(0)},
\end{equation}
where the measurement operators generically read~\cite{Wiseman2009}:
\begin{align} \label{eq:d-operators}
\Omega_{\mathbf{I}(t)} =& \left[\id - i dt H -\frac{dt}{2}\sum_k L_k^\dagger L_k + \sum_k L_k I_k(t)dt \right] \nonumber \\ 
&\times \prod_k \sqrt{p_\mathrm{ost}[I_k(t)]},
\end{align}
and here $p_\mathrm{ost}(I_k) = \sqrt{dt/2 \pi} \exp(-I_k^2/2)$ is the so-called ostensible probability distribution~\cite{Wiseman2009}, ensuring $\int p_\mathrm{ost}(I_k) I_k dt = 0$ and $\int p_\mathrm{ost}(I_k) I_k dt ~ I_l dt = \delta_{k,l} dt$. The above measurement operators obey $\int \Omega_\mathbf{I}^\dagger \Omega_\mathbf{I} dt = \mathds{1}$ and hence the trajectory operators $\pazocal{T}(\gamma_{(0,\tau)})$ lead, as in the quantum-jumps case, to the unconditioned evolution $\rho_\tau = \mathcal{E}(\rho_0)$ when averaging over measurement records.
}

\section{\label{sec:jumps} Thermodynamic Framework}

{In this section we} elaborate on the definition of a general thermodynamic framework within the quantum trajectory formalism. In order to provide a consistent identification of all thermodynamic quantities at the trajectory level, we introduce a two-point measurement scheme (TPM), consisting in the inclusion of projective measurements of arbitrary observables at the beginning and at the end of the indirectly monitored process respectively. This allows us to describe trajectories with fixed end-points. {W}hile their actual implementation {in the laboratory} as projective measurements is not essential{, its inclusion in the definitions is needed to recover a consistent framework}. The TPM scheme has been extensively used in quantum thermodynamics for the derivation of fluctuation theorems~\cite{Esposito2009,Campisi2011,Deffner2011,Sagawa2013} and for the identification of quantum work in the context of energy measurements~\cite{Talkner2007,Talkner2016}. In the following we set up a general thermodynamic formalism combining the TPM scheme and continuous monitoring, {as provided by the quantum trajectory formalism introduced above. We will first formally define trajectories in the scheme and associate to them probabilities, which will be compared to their time-reversed twins. Proceeding in this way will allow us to} introduce some {central} concepts such as microreversibility, local detailed balance, and the entropy flow to the environment, which are key to the later identification of heat, work, and entropy production {in Secs.~\ref{sec:energetics} and \ref{sec:ep}}. 

\subsection{Forward and backward processes in the TPM scheme}

As mentioned above, we assume an initial projective measurement is performed on the density operator of the system $\rho_0$ using a complete set of projectors \{$\Pi_n^0\}_n$, each associated to different eigenspaces (in the most simple case these correspond to rank-1 projectors, associated to pure states $\ket{n}$). This first measurement can be alternatively viewed as a preparation procedure of the system in the eigenspace of a given (arbitrary) observable $O$, {which} verifies by construction $[O,\rho_0]=0$. Let us assume that outcome $n_0$ is obtained (or prepared) in the initial step. After that, the external driving protocol $\Lambda = \{\lambda(t)~;~0\leq t \leq \tau \}$ is executed. The system {continuously monitored until the final time $\tau$} then follows the dynamical evolution dictated by the SSE (or the SME) {with Hamiltonian and Lindblad operators generically depending on the control parameter $\lambda(t)$ at any time}. Along the evolution, the monitoring procedure produces a measurement record ${\gamma}_{(0,\tau)}$. Once the final time $\tau$ is reached, a second projective measurement is performed using another arbitrary complete set of projectors $\{ \Pi_{n}^\tau \}_n$ and outcome $n_\tau$ is obtained. We can therefore define a trajectory $\gamma_{[0,\tau]}$ in this TPM scheme {labeled by a closed interval $[0,\tau]$} as the sequence 
\begin{equation}
\gamma_{[0,\tau]} = \{ n_0, {\gamma}_{(0,\tau)},n_\tau\},    
\end{equation}
which contains both the particular outcomes of the initial and final projective measurements $n_0$ and $n_\tau$, together with the continuous monitoring measurement record ${\gamma}_{(0,\tau)}$. The latter may take different forms depending on the particular measurement scheme chosen for the monitoring (direct detection of jumps, homodyne-like measurement, heterodyne-like measurement, etc) as discussed in Sec.~\ref{sec:quantumtrajectories}.

The probability to observe a given trajectory $\gamma_{[0,\tau]}$ can be decomposed in the probability to sample the initial state $\ket{n_0}$ from $\rho_0$ and to observe a given record ${\gamma}_{(0,\tau)}$ followed by the final outcome $n_\tau$ in the final projective measurement, that is:
\begin{equation} \label{eq:prob}
 P_\Lambda(\gamma_{[0,\tau]}) = p^0_{n_0}  \tr[\Pi_{n_\tau}^\tau \pazocal{T}_\Lambda({\gamma}_{(0,\tau)}) \Pi_{n_0}^0 \pazocal{T}_\Lambda^\dagger({\gamma}_{(0,\tau)})],  
\end{equation}
where $\rho_0 = p_{n}^0~ \Pi_{n}^0$ is the spectral decomposition of the initial density operator. 
The operator $\pazocal{T}_\Lambda({\gamma}_{(0,\tau)})$ generates of the trajectory on the system state associated to the measurement record ${\gamma}_{(0,\tau)}$ {and driving protocol $\Lambda$}. Notice also that we keep the subscript $\Lambda$ in the trajectory probability in Eq.~\eqref{eq:prob} to emphasize that it is conditioned on the external driving protocol.

By averaging over trajectories $\gamma_{[0,\tau]}$ the final state of the system becomes {again} $\rho_\tau = \sum_{n_\tau} \Pi_{n_\tau}^\tau \mathcal{E}(\rho_0) \Pi_{n_\tau}^\tau$, where {we recall that} $\mathcal{E} \equiv {\text{T}_+} \exp({-\int_0^\tau dt\mathcal{L}_{\lambda(t)}})$ is the open-system evolution generated by the master equation~\eqref{eq:Lindblad}. 
Notice that in the special case where the final projective measurement is performed in the density operator eigenbasis (prior to measurement) we have $[\Pi_{n_\tau}^\tau, \mathcal{E}(\rho_0)] = 0$ and hence it does not have any effect at the ensemble level, similarly to what happens with the initial projective measurement. {As mentioned before,} the later situation can be interpreted as if there were not projective measurements in the setup at all, but one is still interested in asking what is the probability of (monitored) two-point trajectories $\gamma_{[0,\tau]}$. Keeping this in mind, we {will} proceed considering the more general setup where the final projectors are arbitrary{, unless otherwise stated}.

\begin{figure*}[tbh]
    \centering
    \includegraphics[width=0.95 \linewidth]{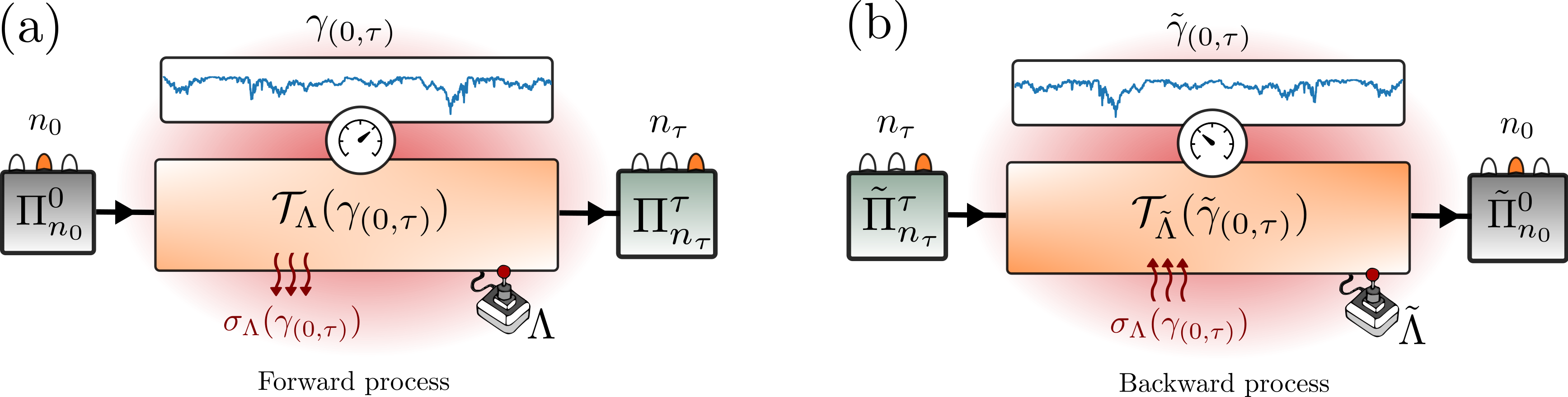}
    \caption{Schematic circuit-like representation of the forward (a) and backward (b) processes when the trajectory $\gamma_{[0,\tau]}=\{n_{0}, \gamma_{(0,\tau)}, n_{\tau} \}$ and its time-reversal twin  $\tilde{\gamma}_{[0,\tau]}=\{n_{\tau}, \tilde{\gamma}_{(0,\tau)}, n_{0} \}$ are respectively recorded (blue traces at the top). In both cases the left and right boxes represent the  preparation/measurement of the system in the corresponding initial and final states, while the central rectangle represents the open-system evolution under monitoring as given by operator $\pazocal{T}_\Lambda(\gamma_{(0,\tau)})$ [ $\pazocal{T}_{\tilde{\Lambda}}(\tilde{\gamma}_{(0,\tau)})$ in the backward process]. External driving is represented by the inclusion of the joystick associated to the execution of the control protocol $\Lambda$ [$\tilde{\Lambda}$ in the backward process] while the monitoring scheme correspond to a diffusive unravelling. Following Eq.~\eqref{eq:ef}, the entropy flow to the environment associated to the record $\sigma_\Lambda(\gamma_{(0,\tau)})$ in the forward process is inverted in the backward process, as illustrated by the red arrows in the bottom of the central rectangles.
    }
    \label{fig:scheme}
\end{figure*}

{In order to assess the reversibility of a dynamical evolution } 
we now compare the process introduced above, which we may refer to as the "forward process", with its time-reversed twin, or "backward process", as defined in the following operational way. Here it is convenient to introduce the time-reversal operator in quantum mechanics, $\Theta$, which is anti-unitary $\Theta \Theta^\dagger = \Theta^\dagger \Theta = \id$ with $\Theta i \id = -i \id \Theta$ and it is the responsible for changing the sign of odd-variables under time-reversal, such a momenta or magnetic fields~\cite{Haake2010}. For the backward process, {it is often convenient}
{to define measurements and dynamics operators; we will distinguish them from the forward ones with a} {tilde}.
{In the backward process} the system is initially prepared in one of the eigenspaces associated to the {corresponding 'reversed'} {version of the}  projectors {used at the end of the forward process}, $\{ \tilde{\Pi}_m^\tau \equiv \Theta \Pi_{m}^\tau \Theta^\dagger \}_n$ (although other choices for the initial state of the backward process are possible in general~\cite{Esposito2009,Campisi2011,Deffner2011} in analogy to the classical case~\cite{Crooks1999,Seifert2005}). Subsequently the time-reversed driving protocol $\tilde{\Lambda} = \{ \lambda(\tau-t) ~;~ 0 \leq t \leq \tau \}$ is implemented over the transformed system Hamiltonian $\tilde{H}(\lambda) \equiv \Theta H(\lambda) \Theta^\dagger$. For example if the system Hamiltonian contains as the only odd variable under time reversal the magnetic field $B$, then the transformed Hamiltonian would be $\tilde{H}(t) = \Theta H(t, B) \Theta^\dagger = H(t, -B)$, see e.g. Ref.~\cite{Andrieux2008}. The system hence evolves under the associated SSE (or SME) associated to such time-inverted driving protocol under continuous monitoring until time $\tau$, where the final projective measurements in performed using the set of projectors $\{ \tilde{\Pi}_n^0 \equiv \Theta \Pi_{n}^0 \Theta^\dagger\}_n$. When the observables measured in the projective measurements are time-reversal invariant, we simply have $\tilde{\Pi}_n^\tau = \Pi_n^\tau$ and $\tilde{\Pi}_n^0 = \Pi_n^0$ for every outcome $n$. In Fig.~\ref{fig:scheme} we schematically illustrate both the forward and the backward processes as introduced above.

We are interested in the case where each of the projective measurements produces the same outcome than in the forward process, and the monitoring procedure registers exactly the time-reversed measurement record $\tilde{\gamma}_{(0,\tau)}$ as compared to the forward record ${\gamma}_{(0,\tau)}$. For the case of quantum jumps this means that $\tilde{\gamma}_{(0,\tau)}=\{(\tilde{k}_J,\tau- t_J), ..., (\tilde{k}_2, \tau - t_2), (\tilde{k}_1, \tau - t_1) \}$ reproduces the inverse sequence of jumps, where $\tilde k_j$ corresponds to the opposite jump ($\tilde{L}_{k_j} \propto L_{k_j}^\dagger$) with respect to $k_j$ in the forward trajectory~\cite{Manzano2018b}. Indeed a jump up in the energy ladder of the system in the forward trajectory corresponds to a jump down when time is reversed, as much as a photon emission in a forward trajectory corresponds to a photon absorption in the time-reversed one. A similar rule applies to the diffusive trajectories scenario for the time-reversed measurement traces of the monitored currents $\tilde{\gamma}_{(0,\tau)}= \{\tilde{{I}}_k(\tau -t) ~;~ 0\leq t \leq \tau\}_{k=1}^K$. Here $\tilde{I}_k$ is the current associated to the adjoint-twin of the Lindblad operator $L_k$ in the set $\{L_k\}_{k=1}^K$. This means that the current generated from monitoring $L_k$ in the forward process, should equal the current generated from the operator $\tilde{L}_k$ in the backward process when the recorded sequence is inverted (although these are obtained with different probabilities), in analogy to the case of the jump trajectories.

{Finally}, combining the time-reversed measurement record with the outcomes of the projective measurements leads to the definition of the time-reversed trajectory $\tilde{\gamma}_{[0,\tau]}=\{n_\tau, \tilde{{\gamma}}_{(0,\tau)}, n_0\}$, which corresponds exactly with the inverse sequence of events than $\gamma_{[0,\tau]}$. The probability of such a time-reversed trajectory hence reads:
\begin{equation} \label{eq:prob-b}
 P_{\tilde{\Lambda}}(\tilde{\gamma}_{[0,\tau]}) = p^\tau_{n_\tau}  \tr[\tilde{\Pi}_{n_0}^0  \pazocal{T}_{\tilde \Lambda}(\tilde{{\gamma}}_{(0,\tau)}) \tilde{\Pi}_{n_\tau}^\tau \pazocal{T}_{\tilde \Lambda}^\dagger(\tilde{{\gamma}}_{(0,\tau)})],  
\end{equation}
where $p_{n_\tau}^\tau= \tr[\Pi_{n_\tau}^\tau \mathcal{E}_\tau(\rho_0)]$ is the probability to sample $n_\tau$ from $\rho_\tau$ at the beginning of the backward process and $\pazocal{T}_{\tilde \Lambda}(\tilde{{\gamma}}_{(0,\tau)})$ is the operator generating the time-reversed record $\tilde{\gamma}_{(0,\tau)}$ under the time-reversed driving protocol $\tilde{\Lambda}$.
When averaging over time-reversed trajectories the final state of the system reads $\tilde{\rho}_\tau = \sum_{n_0}  \tilde{\Pi}_{n_0}^0 \tilde{\mathcal{E}}_\tau(\Theta \rho_\tau \Theta^\dagger) \tilde{\Pi}_{n_0}^0$ where $\tilde{\mathcal{E}}_\tau$ is the quantum Markovian semigroup of the time-reversed dynamics, $\tilde{\mathcal{E}}_\tau \equiv \exp(-\int_0^\tau dt \tilde{\mathcal{L}}_\lambda)$, with generator in Lindblad (GKLS) form:
\begin{equation} \label{eq:backward-Lindblad}
\tilde{\mathcal{L}}_\lambda(\rho) = -i[ \tilde{H}, \rho] + \sum_k \tilde{L}_k \rho \tilde{L}_k^\dagger - \frac{1}{2}\{\tilde{L}_k^\dagger, \tilde{L}_k, \rho \},   
\end{equation}
{where we omitted the dependence on $\lambda$ in $\tilde{H}(\lambda)$ and $\tilde{L}_k(\lambda)$.}
We recall that here $\tilde{H}_0(\lambda) = \Theta H(\lambda) \Theta^\dagger$ and the relation between the Lindblad operators in forward and time-reversed dynamics will be deduced below, following Refs.~\cite{Horowitz2013,Manzano2018b}.

\subsection{Microreversibility and local detailed balance}

The concept of microreversibility, or microscopic reversibility, was originally introduced by Boltzmann in the kinetic theory of gases. It refers to the decomposition of the microscopic dynamical evolution of a system in elementary processes, each of which posses a time-reversal twin~\cite{Boltzmann1964}. The microreversibility principle in quantum mechanics is well-known for closed non-autonomous systems
\begin{equation}\label{eq:U-micro}
\Theta^\dagger U_{\tilde{\Lambda}}^\dagger(\tau-t, \tau) \Theta = U_\Lambda(0,t),    
\end{equation}
where $U_\Lambda(0,t)$ is the unitary evolution of a system under the generic control protocol $\Lambda$, from the initial time up to $t$, and  $U_{\tilde{\Lambda}}^\dagger(\tau-t, \tau)$ the evolution subjected to the time-reversed protocol $\tilde{\Lambda}$, see e.g. Refs.~\cite{Campisi2011,Sagawa2013}. Here we consider its applicability in the case of open quantum systems following quantum trajectories. In particular, it has been proven~\cite{Manzano2018b} (see also Refs.~\cite{Horowitz2012,Horowitz2013}) that, {starting from the the global (system + environment) unitary evolution and applying microreversibility in Eq.~\eqref{eq:U-micro} there}, one can generically relate the trajectory operators in forward and time-reversal dynamics as:
\begin{equation}\label{eq:ef}
\Theta^\dagger  \pazocal{T}_{\tilde \Lambda}^\dagger(\tilde{{\gamma}}_{(0,\tau)}) \Theta =   \pazocal{T}_{\Lambda}({\gamma}_{(0,\tau)}) e^{-\sigma_\Lambda({\gamma}_{(0,\tau)})/2},
\end{equation}
where the {scalar} quantity $\sigma_\Lambda({\gamma}_{(0,\tau)})$ is the stochastic entropy flow from the system to the environment {(or entropy exchange)} accumulated during the trajectory up to time $\tau$ {in $k_B$ units (we will assume $k_B = 1$ thorough)}. {This quantity is often referred to as either (integrated) ``entropy flow"~\cite{VdB2015} or ``entropy of the medium"~\cite{Seifert2012} in stochastic thermodynamics and is associated to the reversible part of the changes in the entropy of the system due to its interaction with the environment~\cite{Prigoginebook}}.
{Notice that since $\sigma_\Lambda$ is not an operator, Eq.~\eqref{eq:ef} is far from trivial}. This relation quantifies in entropic terms the probability associated to the generation of measurement records $\gamma_{(0,\tau)}$ under the driving protocol $\Lambda$ with respect to the time-inverted trajectories $\tilde{\gamma}_{(0,\tau)}$ when the driving protocol is also inverted. The higher the entropy flows to the environment, the less probable is to reproduce the (inverted) forward trajectory in the backward process. Equation~\eqref{eq:ef} is a quantum generalization of the micro-reversibility relation put forward by Crooks in classical systems subjected to {external driving and} thermal fluctuations~\cite{Crooks1999}. {We also note that Eq.~\eqref{eq:ef} reduces to Eq.~\eqref{eq:U-micro} for unitary processes, where $\sigma_\Lambda = 0$.}

Remarkably, the result in Eq.~\eqref{eq:ef} has been derived for generic quantum jump trajectories {and} without further assumptions in the form of the environment, {that} does not need to be thermal~\cite{Manzano2018b}. The only condition is that the \emph{set} of Lindblad (jump) operators $\{L_k\}_{k=1}^K$ is self-adjoint, namely, that the adjoint of every operator $L_k$ is proportional to another operator contained in the set. This condition, together with the {further assumption} that the only operator which commutes with all elements in the set is the identity operator, guarantees that the Lindblad dynamics is equipped with a unique (relaxing) steady-state when the driving is frozen~\cite{Sponh1977,Rivas2012}. {The later is an important condition in the presence of a thermal environment since it enforces that, in absence of driving, the system relaxes back to an equilibrium state.}
However, one may also consider cases where {the stationary state of the system is out of equilibrium, or even when there is more than one invariant state like e.g. in pure decoherence dynamics, for which the second condition above is not necessarily satisfied.}

For an adjoint set $\{L_k\}_{k=1}^K$, the Lindblad (jump) operators are either self-adjoint operators, or come in pairs  $\{L_{k+} ,L_{k-}\}$ such that {$L_{k+} = \sqrt{\Gamma_+} L^\dagger$ and $L_{k-} = \sqrt{\Gamma_-} L$ for some operator $L$}, and verify a generalized local detailed balance relation~\cite{Horowitz2013,Manzano2018b}: 
\begin{equation}\label{eq:ldb}
L_{k+}(\lambda) = L_{k-}^\dagger(\lambda) e^{\Delta s_{k+}(\lambda)/2}
\end{equation}
where $\Delta s_{k\pm}(\lambda) = \pm \log(\Gamma_+(\lambda)/\Gamma_{-}(\lambda))$ {is a real function} and $\Gamma_{\pm}(\lambda)\geq 0$ { represent} the corresponding (time-dependent) rates evaluated at the instantaneous value of the control parameter $\lambda$. For the case of self-adjoint operators we instead have $\Delta s_k(\lambda) = 0$, independently of their rate. These considerations apply to generic environments~\cite{Manzano2018b} composed by several thermal and/or particle reservoirs, or that can be prepared in quantum states, like e.g. the squeezed thermal reservoir~\cite{Manzano2018c} hence generalizing previous results derived for a single thermal reservoir~\cite{Crooks2008,Horowitz2012}, and by assuming a specific form of the system-environment interaction~\cite{Horowitz2013}. The relation \eqref{eq:ldb} hence points to a very fundamental property of Markovian open quantum systems in the weak-coupling limit {as described by GKLS master equations}. {In Sec.~\ref{sec:examples} some simple examples are examined where the operators $\{L_k \}$ correspond either to the spontaneous and stimulated emission and absorption of photons in a thermal electromagnetic environment, or the case in which we have a single self-adjoint operator inducing pure decoherence on the system.}

As commented in the previous section, in the case of quantum jumps the trajectory operator~\eqref{eq:T-operator} consists of a sequence of no-jump evolution operators $\mathcal{U}_\Lambda(t_i, t_j)$ intersected by the instantaneous jumps $L_{k}(\lambda_{t_j})$. The trajectory operator of the time-reversed measurement record $\pazocal{T}_{\tilde{\Lambda}}(\tilde{\gamma}_{(0,\tau)})$ then contains the inverse sequence of operators, but with jumps $\tilde{L}_{{k}_j}(\lambda_{\tau-t_j})$ associated to the paired index $\tilde{k}_j$  (i.e. making the change $k{+} \leftrightarrow k{-}$) and the no-jump evolution operators containing the time-reversal driving $\mathcal{U}_{\tilde{\Lambda}}(\tau - t_i, \tau - t_j)$. Note, however, that $\tilde{L}_{k+} \neq L_{k-}$. To see how Eq.~\eqref{eq:ef} follows from the local detailed balance relation \eqref{eq:ldb}, one inserts pairs $\Theta^\dagger \Theta = \id$ between each the operators inside $\pazocal{T}_{\tilde{\Lambda}}(\tilde{\gamma}_{(0,\tau)})$. Then assuming $\Theta L_k = L_k \Theta$ we obtain from the microreversibility relation for non-autonomous (closed) systems that $\Theta^\dagger \mathcal{U}^\dagger_{\tilde{\Lambda}}(\tau - t_i, \tau - t_j) \Theta = \mathcal{U}_\Lambda (t_i, t_j)$, meaning that the smooth no-jump evolution does not contribute to the entropy flow~\cite{Horowitz2012,Horowitz2013}. Finally we use the local detailed balance relation in Eq.~\eqref{eq:ldb} to convert the backward jumps as 
\begin{equation} \label{eq:backward-jump}
\Theta^\dagger \tilde{L}_{{k}_j}^\dagger(\lambda_{\tau-t_j}) \Theta = L_{k_j}(\lambda_{t_j}) e^{-\Delta s_{k_j}/2}.    
\end{equation}
This leads to recover \eqref{eq:ef} with the accumulated entropy flow during the interval $[0,\tau]$ consisting of a sum of the entropy exchanged with the environment in each jump:
\begin{equation}\label{eq:ef-jumps}
\sigma_\Lambda({\gamma}_{(0,\tau)}) = \sum_{j=1}^J \Delta s_{k_j}(\lambda_{t_j}) = \sum_{k=1}^K  \int_0^\tau dN_k \Delta s_k(\lambda_t), 
\end{equation}
{where in the second equality we have rewritten the expression in terms of the stochastic increments $\{ dN_k\}$ appearing in the SSE (or SME).} Here the interpretation of the entropy flow in terms of the exchange of physical quantities with the system is associated to the basis in which the jump operators $L_{k+}$ and $L_{k-}$ promote the jumps, and that may change during the evolution. 

It is also interesting in many applications to extend the above situation to the case in which the environment is composed by several independent reservoirs, such as thermal reservoirs at different temperatures, or particle reservoirs with different chemical potentials. In such case, the entropy exchanged with the environment can be decomposed as the sum over each reservoir contribution $\sigma_\Lambda({\gamma}_{(0,\tau)}) = \sum_{r=1}^R \sigma^{(r)}_\Lambda({\gamma}_{(0,\tau)})$, by identifying the jumps associated to transitions triggered by the different reservoirs $r = 1, ..., R$, that is:
\begin{equation}
\sigma^{(r)}_\Lambda({\gamma}_{(0,\tau)}) = \sum_{j=1}^{J_r} \Delta s^{(r)}_{k_j}(\lambda_{t_j}) = \sum_{k \in \mathcal{R}_r}  \int_0^\tau dN_k \Delta s^{(r)}_k(\lambda_t),
\end{equation}
where $\Delta s^{(r)}_{k}$ is associated to Lindblad jump operators $L_{k\pm}${from reservoir $r$} following local detailed balance Eq.~\eqref{eq:ldb}, and {we denoted} $J_r$ the total number of jumps triggered by reservoir $r$ during the trajectory $\gamma_{(0,\tau)}$. {In the second equality we also introduced $\mathcal{R}_r$ in the sum to denote the set of channels corresponding to reservoir $r$.}

Performing the average in Eq.~\eqref{eq:ef-jumps} over trajectories $\gamma_{[0,\tau]}$ we obtain:
\begin{align} \label{eq:ef-average}
    \langle \sigma_\Lambda(\gamma_{(0,\tau)}) \rangle_\gamma &= \sum_{\{\gamma_{[0,\tau]}\}} P_\Lambda(\gamma_{[0,\tau]}) \sigma_\Lambda(\gamma_{(0,\tau)}) \nonumber \\ &= \int_0^\tau dt \sum_{k=1}^K \langle L_{k}^\dagger L_k \rangle \Delta s_{k}(\lambda_t),
\end{align}
where in the second line we used the decomposition of the evolution over infinitesimal time-steps introduced in Sec.~\ref{sec:quantumtrajectories}, and the fact that only the jumps contribute to the entropy flow.
We identify the average entropy flow rate as the expression inside the integral above:
\begin{equation}
\langle \dot{\sigma}_\Lambda \rangle_\gamma = \sum_{r=1}^R \langle \dot{\sigma}^{(r)}_\Lambda \rangle_\gamma = \sum_{r=1}^R \sum_{k \in \mathcal{R}_r} \langle L_k^\dagger L_k \rangle \Delta s_{k}^{(r)}(\lambda_t), 
\end{equation}
where in the second equality we split again in the contributions from the different reservoirs. We notice that whenever $\Delta s_k^{(r)} = 0$ for all $k \in \mathcal{R}_r$, the entropy flow to the environment $r$ vanishes. This is the case e.g. for purely decoherence processes associated to self-adjoint Lindblad operators, or to the case of infinite temperature reservoirs, where while leading to both jumps up and down on the energy ladder, $L_{\pm}$, these occur at equal rates $\Gamma_+ = \Gamma_-$ and hence $L^\dagger_{-} = L_{+}$.

\subsection{Microreversibility in diffusive trajectories} \label{subsec:micro-diffusive}

The microreversibility relation in Eq.~\eqref{eq:ef} can be extended to 
diffusive trajectories in some particular cases. Although a derivation for generic situations as in the case of quantum jumps is not available to the best of our knowledge {we provide here some references and hints in this case for the sake of completeness}. One case for which Eq.~\eqref{eq:ef} has been derived (although without including the final projective measurement or end-point of the trajectory) is for a single self-adjoint Lindblad operator describing the monitoring of a system observable~\cite{Auffeves2017}. Microreversibility has been also considered for the case of a two-level (qubit) system monitored under 
Gaussian measurements~\cite{Dressel2017,Manikandan2019b,Manikandan2019,Harrington2019} and the quantity $\sigma_\Lambda$ was identified with a measure of the arrow of time. In these studies the choice of backward operators to time-reverse the trajectory [our operators $\pazocal{T}_{\tilde{\Lambda}}^\dagger(\tilde{\gamma}_{(0,\tau)})$] lacked a proper normalization and hence do not lead in general to a probability distribution, see also Ref.~\cite{Watanabe2014}. This problem disappears in the case in which the Lindblad operators of the monitored system are {all} self-adjoint operators $L_k = L_k^\dagger$ as pointed out in Ref.~\cite{DiStefano2018}. The same issue {arose} previously also for quantum jumps with the similar definitions provided in Ref.~\cite{Leggio2013}, while the proposal in Ref.~\cite{Auffeves2017} run into similar problems when applied to general situations beyond the self-adjoint case.
Examples of diffusive trajectories based on self-adjoint operators in the thermodynamic context have been considered for a two-level system in the context of state-stabilization~\cite{Auffeves2017}, a driven monitored double quantum dot~\cite{Alonso2016}, and the circuit QED setup~\cite{DiStefano2018}. The later approach has been further used to implement a circuit QED Maxwell demon as reported in Ref.~\cite{Naghiloo2018} {(this setup will be examined later in Sec.~\ref{sec:examples})}.

In the following we reproduce Eq.~\eqref{eq:ef} for the case in which all operators used in the unravelling $\{L_k\}_{k=1}^K$ verify $\Delta s_k = 0$ in Eq.~\eqref{eq:ldb}. This includes (but is not restricted to) monitoring system observables, {which corresponds to the case in which all Lindblad operators} are self-adjoint, i.e. $L_k = L_k^\dagger$ for all $k = 1, ..., K$. Since each monitored current in the forward process $\{I_k(t) ~;~ 0 \leq t \leq \tau \}$ is associated to the unravelling of a Lindblad operator $L_k$, we associate the time-reversed current $\{\tilde{I}_k(t) ~;~ 0\leq t \leq \tau \}$ to the unravelling of the corresponding adjoint-twin of the operator, $\tilde{L}_k$. Assuming the same form for the measurement operators than in the forward process at infinitesimal time-steps, Eq.~\eqref{eq:d-operators} in Sec.~\ref{sec:quantumtrajectories}, but replacing the jump operators by their adjoint-twins, we have:
\begin{align} \label{eq:reversed-op}
\tilde{\Omega}_{\mathbf{r}}(\lambda_{\tau - t}) &= \big[\id - i dt \Theta H (\lambda_{\tau -t}) \Theta^\dagger - \frac{dt}{2}\sum_k \tilde{L}_k^\dagger(\lambda_{\tau-t}) \tilde{L}_k(\lambda_{\tau-t})   \nonumber \\ 
&~~+  \sum_k \tilde{L}_k(\lambda_{\tau -t}) \tilde{I}_k(\tau - t)dt \big] \prod_k \sqrt{p_\mathrm{ost}(\tilde{I}_k)}, 
\end{align}
with time-reversed currents $\tilde{I}_k(\tau - t) = I_k(t)$ and ostensible probability ensuring white noise, $\int p_\mathrm{ost}(\tilde{I}_k) \tilde{I}_k dt = 0$ and $\int p_\mathrm{ost}(\tilde{I}_k) (\tilde{I}_k dt)^2 = dt$ as for the forward process. From the above definition, and by using Eq.~\eqref{eq:backward-jump} for $\Delta s_k = 0$, we obtain:
\begin{equation} \label{eq:ldb-diffusive}
\Theta^\dagger \tilde{\Omega}^\dagger_{\mathbf{I}}(\lambda_{\tau-t}) \Theta = \Omega_\mathbf{I}(\lambda_t),    
\end{equation}
where we also used that since the twin operators are also in the set of Lindblad operators, $\sum_k \tilde{L}_k^\dagger \tilde{L}_k = \sum_k {L}_k^\dagger {L}_k$. 
We notice that in the general case (i.e. $\Delta s_k \neq 0$), Eq.~\eqref{eq:ldb-diffusive} is not verified anymore {and microreversibility is lost}. This is due to the structure of the measurement operators in Eq.~\eqref{eq:d-operators}. Whether one can derive more general forms for the measurement operators in the time-reversed process that verify Eq.~\eqref{eq:ef} is an open question that requires further investigation. Nonetheless, it is worth pointing out that any reduced dynamics of the system admits a representation in terms of Kraus operators that verifies Eq.~\eqref{eq:ef}, as explicitly constructed in Ref.~\cite{Manzano2018b}.

Introducing Eq.~\eqref{eq:ldb-diffusive} for each of the measurement operators in the trajectory generator of the backward {process} $\pazocal{T}_{\tilde{\Lambda}}(\tilde{\gamma}_{(0,\tau)}) = \int_{t=0}^\tau dt \tilde{\Omega}_{\mathbf{J}}(\lambda_{\tau - t})$ we recover Eq.~\eqref{eq:ef} with the accumulated entropy flow given by:
\begin{equation} \label{eq:ef-dif}
    \sigma_\Lambda(\gamma_{(0,\tau)}) = 0,
\end{equation}
in agreement with known results for observables monitoring~\cite{DiStefano2018}. This leads to a zero average entropy flow $\langle \sigma_\Lambda(\gamma_{(0,\tau)}) \rangle = 0$, as in the quantum jump trajectory case, c.f. Eq.~\eqref{eq:ef-average}. We remark 
that here, as in the case of quantum jumps, a zero entropy flow is not incompatible with extra entropy production in the environment due to manipulations leading to the monitoring scheme itself.

\section{Energetics and the first law} \label{sec:energetics}

Having introduced the complete setup combining the TPM scheme and continuous monitoring of the system and established the microreversibility relation~\eqref{eq:ef}, we are now in a position to discuss the energetics of the system and formulate the first law of thermodynamics {in a monitored system}. In order to do that, we will first characterize the energy changes during the thermodynamic process at the level of single trajectories and then decompose this quantity into heat and work contributions. Work and heat, contrary to energy changes, are not state functions, i.e. they depend on the precise path the system follows during the evolution, and have been often associated to ordered (controllable) and disordered (uncontrollable) forms of energy. We anticipate that this distinction should not be done in an arbitrary way, only based on the subjective view of which part of the energy is useful or not for one's a priori purposes. On the contrary, in a consistent thermodynamic approach the disordered character of the energy currents needs to be founded in the reversibility/irreversibility properties of the evolution in connection with the environment, or said in another way, in the fundamental link between energy and entropy.

We start by introducing the expected energy of the system conditioned on the measurement outcomes at any time as $E_\gamma(t) = \tr[H(\lambda_t) \rho_\gamma(t)]$, where $\rho_\gamma(t)$ denotes the state of the system conditioned on $\gamma_{[0,\tau]}$ {and $H(\lambda_t)$ is the inclusive Hamiltonian of the system including driving contributions}. Following this definition, the energy change in the system along the whole trajectory $\gamma_[0,\tau]$ is:
\begin{align} \label{eq:energy}
\Delta E_\Lambda(\gamma_{[0,\tau]}) &= E_\gamma(\tau) - E_\gamma(0) \nonumber \\
&=~ \tr[H(\lambda_\tau) \Pi_{n_\tau}^\tau] - \tr[H(\lambda_0) \Pi_{n_0}^0],
\end{align}
which only depends on the initial and final states of the trajectory as given by the projectors $\Pi_{n_0}^0$ and $\Pi_{n_\tau}^\tau$ respectively, that is, the energy changes is a state function. Notice that Eq.~\eqref{eq:energy} is still in general a difference between expected values for the system energy, and not a difference between energy eigenstates like in the energetic TPM~\cite{Campisi2011}. That limit is recovered in the specific case for which the system starts in a diagonal state in the initial Hamiltonian basis and the final Hamiltonian is measured at the end. From the energy change values~\eqref{eq:energy} one can formally construct a probability distribution for the energy changes along the process as 
\begin{equation}
P_\Lambda(\Delta E) = \sum_{\{ \gamma_{[0,\tau]}\}} P_\Lambda(\gamma_{[0,\tau]}) \delta (\Delta E - \Delta E_\Lambda(\gamma_{[0,\tau]})),    
\end{equation}
where $\Delta E_\Lambda(\gamma_{[0,\tau]})$ is given by Eq.~\eqref{eq:energy}{, $P_\Lambda(\gamma_{[0,\tau]})$ in Eq.~\eqref{eq:prob}}, and $\delta$ denotes the Dirac delta.

\subsection{Stochastic Heat} \label{subsec:heat}

In order to characterize heat in a generic process, we will need to further specify the environment that produces the open system evolution and its effect on the system dynamics. Let us assume that the environment is composed by a set of reservoirs labeled by the index $r= 1 ... R$ that are able to exchange energy or other globally conserved quantities (also called charges) with the {system}. 
{We notice that treating the case of several conserved quantities we expand the scope of the review. }
The reservoirs can then be characterized by a set of temperatures $\{ T_r\}$ and eventually a set of extra potentials $\{ \mu_r^i\}$ associated to conserved quantities. We define the stochastic heat transferred from each reservoir to the system during the trajectory $\gamma_{[0,\tau]}$ as:
\begin{align} \label{eq:heat}
Q_{\Lambda}^{(r)}(\gamma_{(0,\tau)}) &= - T_r ~ \sigma_\Lambda^{(r)}(\gamma_{(0,\tau)}) \nonumber \\
&= -  T_r \sum_{k \in \mathcal{R}_r}  \int_0^\tau dN_k \Delta s^{(r)}_k(\lambda_t),    
\end{align}
where $\sigma_\Lambda^r(\gamma_{(0,\tau)})$ is the entropy flux to the reservoir $r$ as discussed in section ~\ref{sec:jumps} from the microreversibility relation~\eqref{eq:ef}. We can hence define the probability distribution of the heat to reservoir $r$ during the trajectory as:
\begin{equation}
P_\Lambda(Q_r) = \sum_{\{ \gamma\}} P_\Lambda(\gamma_{[0,\tau]}) \delta(Q_r +  T_r \sigma_\Lambda^{(r)}(\gamma_{(0,\tau)})).
\end{equation}
Although Eq.~\eqref{eq:heat} may seem an abstract identification at first sight, its meaning is clarified as soon as we make some extra assumptions that allow us to identify the heat above with the exchange of physical quantities between the system and reservoirs. Note also that from Eq.~\eqref{eq:ef-dif}, the heat transferred to the environment in the {(microreversible)} diffusive trajectories introduced above is zero. 

Let's first consider the case in which the reservoir $r$ is a thermal reservoir at temperature $T_r$, {with state} $\rho_r = \exp(-H_r/T_r)/Z_r$, $H_r$ {being} the reservoir Hamiltonian, and $Z_r = \tr[\exp(-H_r/T_r)]$ the partition function. It exchanges energy with the system through the jump operators $\{L_k\}_{k \in \mathcal{R}_r}$ in the system energy basis, with associated energy quanta $\Delta E_k$ representing the line-width of the transitions. Here and in the following, for the ease of notation, we not always indicate explicitly the dependence of the Hamiltonian or of the operators $L_k$ with the control parameter $\lambda$. However, it should be intended that such dependence may always exists. {We define a ``bare'' or ``effective'' system} Hamiltonian $H_S(\lambda_t)$, {as the piece of the inclusive Hamiltonian, $H = H_S + V$, verifying}: 
\begin{equation}\label{eq:energy-jumps}
 [H_S, L_{k}] = - \Delta E_k L_k, ~~~~\mathcal{L}_\lambda \left(\frac{e^{-\frac{H_S}{T_r}}}{Z} \right) = 0, 
\end{equation}
where $\mathcal{L}_\lambda^{(r)}$ is the piece of the Lindbladian associated to the reservoir $r$ and $Z=\tr[e^{-H_S/T_r}]$. That is, $H_S$ is the piece of the system Hamiltonian that determine the basis in which energy is exchanged with the reservoirs. We note that it may or may not coincide with $H$ generating the unitary part of the dynamical evolution in the Lindblad master equation, Eq.~\eqref{eq:Lindblad}. For example, when the driving is weak, $H_S$ represents the bare Hamiltonian of the system not including the driving contribution {$V$, which is treated as a small perturbation of $H_S$}.
This is the case e.g. {for a two-level system coherently driven by a resonant field~\cite{Alonso2016,DiStefano2018,Naghiloo2018}, as considered in the examples of Sec.~\ref{sec:examples}}, or for the periodic driving of a cavity mode reported in Ref.~\cite{Manzano2018b}. {Another example is compound systems dissipating into local baths coupled by a weak interaction among them, in which case $H_S$ represents the Hamiltonian of the uncoupled systems and $V$ their (weak) interaction Hamiltonian. On the other hand for strong driving or strongly interacting compound systems we generically have $H_S = H$ (and hence $V=0$).} The entropy changes in Eq.~\eqref{eq:ldb} are hence in all these cases:
\begin{equation}
 \Delta s_k^{(r)}(\lambda_t) = - \frac{\Delta E_k(\lambda_t)}{T_r},
\end{equation}
which can be seen as a consequence of the local detailed balance condition for the rates, ensured by the Kubo-Martin-Schwinger (KMS) condition for the reservoir state~\cite{Spohn1978}.

Let us now extend the situation to energy and particles reservoirs, $\rho_r = \exp(-(H_r-\mu_r N_r)/T_r)/Z_r$ with $N_r$ the number of particles operator, $\mu_r$ the corresponding chemical potential, and as before $Z_r = \tr[\exp(-(H_r-\mu_r N_r)/T_r)]$. Assuming that the system exchanges simultaneously both energy and particles with the corresponding reservoir $r$ through the jump operators $\{L_k\}$ with ${k \ in \mathcal{R}_r}$, and denoting $N_S(\lambda_t)$ the number of particles operator in the system (with $[H_S,N_S] = 0$) we have both:
\begin{eqnarray}
[H_S, L_k] = -\Delta E_k L_k^{(r)} ~~;~~ [N_S, L_k] = -\Delta N_k L_k,
\end{eqnarray}
leading to a ``local'' {(e.g. for a single reservoir)} steady state $\mathcal{L}_\lambda^{(r)} (\exp[-(H_S-\mu_r N_S)/T_r]/Z) = 0$. The entropy changes can then be identified following similar lines as:
\begin{equation}
\Delta s_k^{(r)}(\lambda_t) = - \left( \frac{\Delta E_k(\lambda_t)}{T_r} - \mu_r \frac{\Delta N_k(\lambda_t)}{T_r} \right).
\end{equation}
In many applications such as in setups considering quantum dots coupled to electronic reservoirs, these relations can be further simplified from the so-called tight-coupling condition. The tight-coupling condition establish the proportionality of energy and particle currents as $\Delta E_k = \epsilon_k \Delta N_k$ for some parameters $\epsilon_k$, see e.g. the reviews in  Refs.~\cite{Esposito2009,Benenti2017} for relevant examples.

The above relations can be extended to generic situations where the reservoir $r$ is in a generalized Gibbs ensemble and exchanges a set of globally conserved charges with the system $\{X_i\}$ for $i= 1, ..., N$, where $X_i(\lambda_t)$ are Hermitian system operators that may also depend on the external control variable $\lambda_t$. Apart from energy and particles, other extra charges may even not commute with the Hamiltonian, like the different components of angular momenta~\cite{Vaccaro2011,Vaccaro2017,Popescu2020}, or the squeezing asymmetry operator in bosonic fields~\cite{Manzano2018d,Manzano2020}. In such generic cases it is not possible in general to associate the Lindblad jumps operators to the exchange of a single conserved quantity. Nevertheless we can always associate the entropy changes in Eq.~\eqref{eq:ldb} to the collective exchange of charges as: 
\begin{equation} \label{eq:ladder-ops}
\sum_{i=1}^N \frac{\nu_{i}^{(r)}}{T_r} [ X_i,L_{k}] = \Delta s^{(r)}_{k} L_{k}, ~~~~ \mathcal{L}_\lambda^{(r)} \left(\frac{e^{-\sum_i \frac{\nu_{i}^{(r)}}{T_r} X_i}}{Z}\right) = 0.
\end{equation}
Here the set $\{ \nu_{i}^{(r)} \}_{i=1}^N$ are generalized chemical potentials associated to the $N$ conserved quantities in the setup {(and reservoir $r$)}. Only in the case in which all conserved quantities commute we will have $[ X_i,L_{k}] = \Delta X_{k, i} L_{k}$ allowing the split of the entropy changes:
\begin{equation} \label{eq:generic}
\Delta s^{(r)}_{k}(\lambda_t) = \sum_{i=1}^N  \nu_{i}^{(r)}~ \frac{\Delta X_{k, i}(\lambda_t)}{T_r},   
\end{equation}
in the contributions from the stochastic changes $\Delta X_{k, i}$ associated to the transmission of a quantum of the corresponding charge $i$ to the reservoir $r$ with generalized potential $\nu_{i}^{(r)}$. Notice that above we {would} have $\nu_{i}^{(r)} = -1$ for energy jumps and $\nu_{i}^{(r)} = \mu_r$ for particle jumps. In {the commuting} case the heat current to reservoir $r$ is decomposed in the empirical currents to that reservoir as
\begin{equation} \label{eq:heat-commuting}
\dot{Q}^{(r)}_\Lambda(\lambda_t) = \sum_{i=1}^N  \nu_{i}^{(r)}  ~\dot{X}_{k}(\lambda_t),
\end{equation}
with $\dot{X}_k(\lambda_t) \equiv (dN_k/dt) ~\Delta X_{k,i}(\lambda_t)$ and we recall that $\{dN_k\}$ are the stochastic increments of the SSE (or SME) associated to detections on reservoir $r$. The average heat current from reservoir $r$ reads in general:
\begin{equation} \label{eq:av-heat}
 \langle \dot{Q}_\Lambda^{(r)}(t) \rangle_\gamma = \sum_{k \in \mathcal{R}_r} \sum_{i=1}^N \nu_{i}^{(r)} ~\tr[X_i \mathds{D}_k(\rho_t)],   
\end{equation}
which follows from Eq.~\eqref{eq:ef-average} upon using the commutation relations in Eq. ~\eqref{eq:ladder-ops}. Notice that we have contributions to the heat current from every conserved quantity, weighed by their corresponding generalized potentials. For energy exchange the above equation reduces to the prototypical expression $\langle \dot{Q}_\Lambda^{(r)}(t) \rangle_\gamma = \sum_{k \in \mathcal{R}_r} \tr[H_S \mathds{D}_k^{(r)}(\rho_t)]$ known in standard scenarios~\cite{Alicki1979,Kosloff2014} {(for a discussion of local dissipation scenarios see Ref.~\cite{Imparato2021})}. 

In summary, the above definition of heat from the entropy flow in Eq.~\eqref{eq:heat} is a fundamental identification of the ``uncontrollable'' part of the system energy changes (as well as the changes in other globally conserved quantities) as those that are able to increase or decrease the entropy of the reservoir involved in such exchanges. This is in agreement with the classical notion of heat also used in stochastic thermodynamics~\cite{Seifert2012,VdB2015} or in full counting statistics~\cite{Esposito2009} {to describe fluctuations in energy and particle currents}. The above characterization of heat along trajectories follows and extends the one in Refs.~\cite{Horowitz2012,Hekking2013,Suomela2015,Campisi2015,Manzano2015} for thermal reservoirs. We also notice that our notion of heat has been referred to as ``classical heat'' in Refs.~\cite{Auffeves2017,Elouard2017b}. 
However it is worth remarking that it will capture quantum effects as soon as there are different charges that do not commute with each other, $[X_i, X_j]\neq 0$ for some $i,j$, for which the heat exchange cannot be decomposed in separate contributions.

\subsection{Stochastic Work} \label{subsec:work}

Due to the presence of quantum effects, work becomes a subtle concept in quantum thermodynamics when it comes to fluctuations and its characterization has lead to a number of debates about different approaches that one may follow, see e.g. Refs.~\cite{Allahverdyan2005,Talkner2007,Allahverdyan2014,Talkner2016,Deffner2016,Hofer2016,Perarnau2017}. Within a closed energy TPM scheme, however, work can be satisfactorily determined from the outcomes of initial and final measurements whenever the initial measurement do not disturb, on average, the system state. This setup has been indeed used to obtain work fluctuation theorems in general closed and open systems alike~\cite{Campisi2011} {(but with the inconvinience that one would need to perform projective measurements in the whole environment)}. In the present situation it is worth recalling that our TPM does not determine necessarily energy eigenstates of the system and the environment at the initial nor the final points of the trajectory, and hence the situation becomes more tricky.

A simple way to avoid difficulties in the characterization of work consists in \emph{assuming} the verification of the first law, so that work is defined as the deficit between the changes in energy during the trajectory and the heat as identified above:   
\begin{equation} \label{eq:work}
    W_\Lambda(\gamma_{[0,\tau]}) = \Delta E(\gamma_{[0,\tau]}) - Q_\Lambda(\gamma_{[0,\tau]}).
\end{equation}
This is the approach generically adopted following the identification of heat above~\cite{Horowitz2012,Hekking2013,Suomela2015,Campisi2015,Manzano2015,Miller2021}. However, it was also noticed as early as in the inception of this approach in Ref.~\cite{Horowitz2012}, that this definition of work cannot be entirely ascribed to the mechanical work associated to the execution of the driving protocol, the latter being
\begin{equation}\label{eq:drive-work}
 {W}_\Lambda^{\mathrm{drive}}(\gamma_{[0,\tau]}) = \int_0^\tau dt \tr[\dot{H}(\lambda_t) \rho_\gamma(t)].
\end{equation}
This issue emerges even in the case of a single thermal bath, in stark contrast with {classical} stochastic thermodynamics. In the following, we will keep the definition in Eq.~\eqref{eq:work} and decompose $W_\Lambda(\gamma_{[0,\tau]})$ in order to illustrate how all contributions look like in a general setup. In the derivation we will also discuss some other interpretations given in the literature to different contributions arising inside ${W}_\Lambda(\gamma_{[0,\tau]})$. 

In order to proceed 
{we assume the work in} Eq.~\eqref{eq:work}, {and} consider the instantaneous energy changes in the expected energy of the system, that is
\begin{equation}\label{eq:currents}
    \dot{E}_\Lambda(t) = \dot{W}_\Lambda^{\mathrm{drive}}(t) + \tr[H(\lambda_t) \dot{\rho}_\gamma(t)],
\end{equation}
{where we already identified the driving work in Eq.~\eqref{eq:drive-work}}.
{Some previous works proposed to identify the second term in the above equation as a heat current~\cite{Auffeves2017,Alonso2016}, in analogy to standard master equation situations for systems exchanging energy with thermal reservoirs in the weak coupling limit, as elaborated e.g. in Ref.~\cite{Alicki1979}.}
Here we refrain to identify the whole second term in the above equation as a heat current, since such reasoning is not correct in {more general} cases e.g. when extra conserved quantities arise (for example  particle exchange). Henceforth, {we {will not} assume} such an {\emph{a priori}} identification, {but elaborate the distinction between heat and work from the relation of energy currents with the entropy flow between system and environment} (we will turn back to this point later). 

Let us now focus in the case of quantum jump trajectories{, Eq.~\eqref{master_jumps}}. We can decompose the second term in Eq.~\eqref{eq:currents} in drift and jumps contributions as: 
\begin{equation} \label{eq:intermediate}
 \tr[H d{\rho}_\gamma] = -{dt} \sum_k \tr[H \mathds{M}_k(\rho_\gamma)] + \sum_k dN_k \tr[H \mathds{J}_k(\rho_\gamma)], 
\end{equation}
where $\tr[H \mathds{M}_k(\rho_\gamma)] = \tr[H \{L_k^\dagger L_k, \rho_\gamma\}]{/2} - E_\gamma \tr[ L_k^\dagger L_k \rho_\gamma]$ is non-zero when $\rho_\gamma$ shows coherence in the energy basis, and the second term $\tr[H \mathds{J}_k(\rho_\gamma)] = \tr[H L_k \rho_\gamma L_k^\dagger]/\tr[L_k^\dagger L_k \rho_\gamma] - E_\gamma$ accounts for the change in the energy of the system during a jump.
In order to obtain work contributions from Eq.~\eqref{eq:intermediate}, we can subtract from it the infinitesimal heat increments from each reservoir $dQ^{(r)}_\Lambda(t) = - T_r \sum_k dN_k(t) \Delta s_k(\lambda_t)$ as follows from Eq.~\eqref{eq:heat}. This will give us {two} new contributions to work apart from the driving work identified above. Assuming a set of conserved quantities $\{X_i\}_{i=1}^N$ including energy exchange ($X_1= H$), we obtain the three following components for the power performed over the monitored system:
\begin{equation} \label{eq:work-split}
   \dot{W}_\Lambda(t) = \dot{W}_\Lambda^{\mathrm{drive}}(t) + \dot{W}_\Lambda^{\mathrm{chem}}(t) + \dot{W}_\Lambda^{\mathrm{meas}}(t), 
\end{equation}
which correspond to driving, chemical power induced by the extra conserved quantities, and a measurement contribution from the monitoring scheme with zero average. In the following, we give details and provide pertinent comments on the three contributions {(see Appendix~\ref{app1} for details on the derivation and the more general case where $X_1 = H_S \neq H$)}.

The driving power $\dot{W}_\Lambda^{\mathrm{drive}}(t) = \tr[\dot{H}(\lambda_t) \rho_\gamma]$ was already  introduced in Eq.~\eqref{eq:drive-work} and accounts for both the modulation of the energy levels of the system and the coherent evolution of the energy eigenstates. Its average over trajectories reads  $\langle \dot{W}_\Lambda^{\mathrm{drive}}(t) \rangle = \tr[\dot{H}(\lambda_t) \rho(t)]$, which reproduces the standard identification of driving power in weak-coupling thermodynamics with thermal reservoirs~\cite{Alicki1979, Kosloff2014}.

In the second contribution in Eq.~\eqref{eq:work-split} we identified the chemical work performed by the reservoirs associated to the {extra} charges $i > 1$ as:
\begin{equation} \label{eq:work-chemical}
\dot{W}_\Lambda^{\mathrm{chem}}(t) = \sum_{i=2}^N \sum_{r=1}^R \sum_{k \in \mathcal{R}_r} \frac{dN_k}{dt}~ \nu_i^{(r)}~ \frac{\tr[ X_i  \mathds{D}_k (\rho_\gamma)]}{ \langle L_k^\dagger L_k\rangle}
\end{equation}
We notice that $\dot{W}_\Lambda^{\mathrm{chem}}(t)$ cannot be associated to a particular reservoir, but it is a collective contribution, as corresponds to work~\cite{Manzano2020b}. Its average over trajectories simply reads: 
\begin{equation}
\langle \dot{W}_\Lambda^{\mathrm{chem}}(t) \rangle_\gamma = \sum_{i=2}^N \sum_{r=1}^R \sum_{k\in \mathcal{R}_r} \nu_i^{(r)} \tr[X_i \mathds{D}_k(\rho)].    
\end{equation}
For the case of commuting charges, $[X_i, X_j] = 0$ for all $i,j$ the above expression~\eqref{eq:work-chemical} simplifies{, according to Eq.~\eqref{eq:heat-commuting}}, to: 
\begin{equation} \label{eq:work-ch-commuting}
\dot{W}_{\Lambda}^{\mathrm{chem}}(t) = \sum_{i=2}^N \sum_{r=1}^R \sum_{k\in \mathcal{R}_r} \frac{dN_k}{dt}~ \nu_i^{(r)}~\Delta X_{k,i}(\lambda_t)
\end{equation}
with $\Delta X_{k,i}$ the changes in each extra conserved quantity $i>1$ as introduced in Eq.~\eqref{eq:generic}. For example for particle exchange, $\Delta X_{k,i} = \Delta N_k$ {represent the number of particles entering the reservoirs} and $\nu_i^{(r)} = \mu_r$ their chemical potentials.  Henceforth Eq.~\eqref{eq:work-ch-commuting} reduces to the standard chemical work~\cite{VdB2015}.
Notice that, in any case, this contribution to the work is proportional to the stochastic increments  $dN_k(t) =\{ 0, 1\}$ and is hence {of purely stochastic nature} (and associated to the quantum jumps). It accounts for work contributions such as electric currents triggered by the exchange of electrons with {metallic leads acting as} the reservoirs. 

\subsection{Measurement work vs. quantum heat}

Following the above derivation, the third contribution in Eq.~\eqref{eq:work-split} corresponds to the work performed by the continuous measurement process:
\begin{align} \label{eq:meas-work}
 \dot{W}_\Lambda^{\mathrm{meas}}(t) &= - \sum_k \tr[H \mathds{M}_k(\rho_\gamma)] \nonumber \\ 
 &~+\sum_k \frac{dN_k}{dt}~ \left(\frac{1}{2} \frac{\tr[\{H, L_k^\dagger L_k \} \rho_\gamma]}{\langle L_k^\dagger L_k \rangle} - E_\gamma \right),
\end{align}
which shows terms associated to both the drift periods of the evolution and the jumps.
This contribution to the work introduces extra fluctuations during the trajectories but its average vanishes since {these two} terms compensate each other:
\begin{equation}
 \langle \dot{W}_\Lambda^{\mathrm{meas}}(t) \rangle_\gamma = 0 ~~\forall t.
\end{equation}
The fluctuations in Eq.~\eqref{eq:meas-work} are however, non-zero in general. They vanish only if the monitored system $\rho_\gamma$ is maintained in either an eigenstate of $H(\lambda)$, or in a eigenstate of the operators $L_k^\dagger(\lambda) L_k(\lambda)$ for all $k$ during the whole trajectory, as it happens in the classical case. 

{In order to derive Eq.~\eqref{eq:meas-work} we assumed for simplicity that the entire system Hamitlonian is a conserved quantity between system and environment, i.e. within the set of conserved charges $\{ X_i\}$ we took $X_1 = H$. In appendix \ref{app1} we show that in the case of local energy conservation only in part of the system Hamiltonian, that is $X_1 = H_S$ with $H = H_S + V$ (and $V$ a weak perturbation) the expression in Eq.~\eqref{eq:meas-work} is also recovered by simply replacing $H$ by $H_S$, and adding the extra term in Eq.~\eqref{eq:meas-work-interaction} that accounts for (zero-average) extra fluctuations induced by the perturbation $V$.}

The appearance of the quantity~\eqref{eq:meas-work} in the energy balance has been first noticed in Ref.~\cite{Horowitz2012} for a dragged dissipative harmonic oscillator and further explored in Ref.~\cite{Manzano2015} for more general situations. Moreover, it was shown to be a crucial term for recovering work fluctuation theorems along trajectories~\cite{Horowitz2012,Manzano2015}, in line with previous works on the role of projective measurements on the work distribution~\cite{Campisi2010,Campisi2011b}. {This} quantity was also studied in Ref.~\cite{Auffeves2017}, where it was named ``quantum heat'', and further considered in several examples in following works~\cite{Elouard2017b,Gherardini2018b,Mohammady2020}. While there is not a definitive consensus on the status of the energy contribution in Eq.~\eqref{eq:meas-work} different arguments in favor of considering it as either work or heat have been given. In favor of calling it quantum heat, it has been argued that it is of stochastic nature, and hence similar to heat exchanges, in contrast to the work exerted by an external coherent field~\cite{Auffeves2017,Elouard2018}. In favor of the identification as measurement work that we follow here, we have seen that these fluctuations are not related to the exchange of entropy with the environment, which is a fundamental characteristic of work in contrast to heat. In this sense this {stochastic} work contribution would be analogous to the work produced by electric currents~\cite{Benenti2017}, noisy non-conservative forces~\cite{Levy2012}, or thermal reservoirs at infinite temperature~\cite{Levy2012,Correa2014,Strasberg2017}, all of which are also of stochastic nature.

{We also stress} that the quantity in Eq.~\eqref{eq:meas-work} is still an expectation value over the conditional state $\rho_\gamma(t)$ which is not directly associated to the result of any energy measurements. In other words, while this energy change is related to the update of our knowledge about the system's energy during the evolution due to the monitoring process, it does not necessary correspond to any ``real'' (measurable) energy exchange with the measurement apparatus, contrary to the quantities $\Delta E_k$ and $\Delta X_{k,i}$ above. In this sense another {possible} interpretation would be to not identify $ \dot{W}_\Lambda^{\mathrm{meas}}(t)$ with any energy exchange {i.e. refraining from making the assumption in Eq.~\eqref{eq:work} in the first place.}. A closely related interpretation in the context of a quantum heat engines was recently put forward which describes probabilistic violations of the first law~\cite{Kerremans2021}.

For diffusive trajectories, a similar derivation of the work components applies for the cases considered in Sec.~\ref{subsec:micro-diffusive}, that is, monitoring of system observables or more generally, when the set of Lindblad operators is such that $\Delta s_k = 0$ for all $k$. In the present case the stochastic heat is zero according to Eq.~\eqref{eq:heat}. We obtain again the three components to the stochastic power in Eq.~\eqref{eq:work-split}, where now:
\begin{align}\label{eq:meas-work-diffusive}
\dot{W}_\Lambda^{\mathrm{\mathrm{chem}}}(t) =&  \sum_{r=1}^R \sum_{k \in \mathcal{R}_r} \nu_i^{(r)}~ \tr[ X_i  \mathds{D}_k (\rho_\gamma)], \nonumber \\
\dot{W}_\Lambda^{\mathrm{meas}}(t) =& \sum_k \xi_k \left(\langle H L_k \rangle - E_\gamma \langle L_k \rangle \right) + \mathrm{h.c.} 
\end{align}
where $\xi_k = dw_k/dt$ are white noise contributions from the Wiener increments with zero mean and $\langle \xi_k^\ast \xi_l \rangle_\gamma = \delta_{k,l}$. Since taking the average over trajectories we have $\langle \xi_k \rangle_\gamma = 0$ we again obtain zero average measurement work, $\langle \dot{W}_\Lambda^{\mathrm{meas}}(t) \rangle_\gamma = 0$. 
The same arguments as for the case of quantum jump trajectories apply also here, while we recall that 
in this case either the Lindblad operators are {assumed to be} self-adjoint or {that} we have equal rates for complementary jumps.

{Finally, we notice that an extra contribution to the total work $W(\gamma_{[0,\tau]})$ in Eq.~\eqref{eq:work}, comes from the final measurement implemented in the TPM scheme. We include it into the measurement work as
\begin{equation}
W_\Lambda^\mathrm{meas}(\gamma_{[0,\tau]}) = \int_0^\tau \dot{W}_\Lambda^\mathrm{meas}(t) + W_\mathrm{TPM}(\gamma_{[0,\tau]}),
\end{equation}
where the second term due to the projective measurement simply reads
\begin{eqnarray}
W_\mathrm{TPM}(\gamma_{[0,\tau]}) = \tr[H(\lambda_\tau) \Pi_m^{\tau}] - \tr[H(\lambda_\tau) \rho_\gamma(\tau)].
\end{eqnarray}
This contribution has an average $\langle W_\mathrm{TPM}(\gamma_{[0,\tau]}) \rangle_\gamma = \tr[H(\lambda_\tau) [\rho_\tau(\tau) - \mathcal{E}(\rho_0)]]$ which becomes zero whenever the final measurement of the TPM scheme is either an energy measurement ($[H(\lambda_\tau), \Pi_m^\tau] = 0$) or when it is performed in the eigenbasis of the (average) system state $[\mathcal{E}(\rho_0), \Pi_m^\tau] = 0$, in which case $\rho_\tau(\tau) = \mathcal{E}(\rho_0)$.}

\section{Entropy production and Irreversibility} \label{sec:ep}

The second law of thermodynamics establishes that the changes in the entropy of the universe due to an irreversible process are positive. Such statement of the second law is valid in macroscopic thermodynamics, but becomes blurred in the microscopic world, where it is only verified on average~\cite{Jarzynski2011}. Nevertheless it is possible to introduce microscopic quantifiers of irreversibility such as the stochastic entropy production used in stochastic thermodynamics~\cite{Seifert2012,VdB2015}. In the following we show how the notion of stochastic entropy production can be extended to quantum trajectories in the present framework. Then show how general fluctuation theorems can be obtained using this notion and discuss its split into adiabatic and non-adiabatic components, accounting for different sources of irreversibility.

\subsection{Stochastic entropy production}

Quantum versions of the stochastic entropy production have been indeed introduced in TPM schemes~\cite{Deffner2011} and extended to quantum jump trajectories~\cite{Horowitz2013} and to general CPTP maps~\cite{Manzano2015,Manzano2018b} (for a recent review on the subject see Ref.~\cite{Landi2021}).
The two main ingredients for the construction of stochastic entropy production along quantum trajectories  are the identification of a stochastic entropy associated to the system state along a trajectory, and the entropy flux transferred to the environment. We identify the changes in the entropy of the system along a quantum trajectory $\gamma_{[0,\tau]}$ as:
\begin{equation} \label{eq:sys-entropy}
 \Delta S(\gamma_{[0,\tau]}) = -  \log p_{n_\tau}^\tau + \log p_{n_0}^0,
\end{equation}
which correspond to the changes in Shannon {self-}information or surprisal of the system between the initial and final end-points, as in the classical case~\cite{Crooks1999,Seifert2005}. We notice that for being the quantity $\Delta S(\gamma_{[0,\tau]})$ well-defined, the initial and end-points of the trajectories as specified in the TPM are needed. This is also a requisite in order to recover the changes in the von Neumann entropy of the system when the average {over trajectories} is performed:
\begin{eqnarray} \label{eq:vN}
\langle \Delta S (\gamma_{[0,\tau]})\rangle_\gamma = S_\mathrm{vN}(\rho_\tau) - S_\mathrm{vN}(\rho_0),
\end{eqnarray}
with $S_\mathrm{vN}(\rho) = - \tr[\rho \log \rho]${, and $\rho_\tau$ being the system state after the final measurement.} We notice that related proposals not using a TPM~\cite{Leggio2013,Auffeves2017,Elouard2017b} {unfortunately} do not verify, in general, Eq.~\eqref{eq:vN}, hence leading to non-standard entropy production definitions.

The microreversibility relation for the trajectory operators in Eq.~\eqref{eq:ef} implies that the entropy flow to the environment is related to the conditional probabilities of observing the forward and time-reversed trajectories:
\begin{eqnarray} \label{eq:ef-probs}
\sigma_\Lambda (\gamma_{(0,\tau)}) = \log{ \left(\frac{p_\Lambda[n(\tau), {\gamma}_{(0,\tau)} ~|~ n(0)]}{p_{\tilde{\Lambda}}[n(0), \tilde{{\gamma}}_{(0,\tau)} ~|~ n(\tau)]} \right)},
\end{eqnarray}
which have been shown [according to Eq.~\eqref{eq:ef}] to be independent of the initial and final outcomes of the TPM, $n(0)$ and $n(\tau)$, unlike the forward and backward conditional probabilities itself, c.f. Eqs.~\eqref{eq:prob} and \eqref{eq:prob-b} respectively. The entropy flow has been related with the changes in the entropy of the reservoir(s) by analyzing the underlying generalized measurement scheme at each infinitesimal time-step of the quantum trajectory dynamics~\cite{Manzano2018b} (see also Ref.~\cite{Elouard2018}). 
Such analysis leads to the identification of the entropy flow with the von Neumann entropy change of the reservoir due to the interaction with the system plus an extra non-negative term accounting for the internal relaxation of the reservoir back to its equilibrium state: 
\begin{equation}
\langle {\sigma}_\Lambda^{(r)} \rangle_\gamma = S(\rho_r^\prime) - S(\rho_r) + S(\rho_r^\prime || \rho_r), 
\end{equation}
where we denoted $\rho_r$ and $\rho_r^\prime$ the density operator of the reservoir $r$ before and after the evolution step from $t$ to $t+dt$ respectively, and $S(\rho_A || \rho_B) = \tr[\rho_A (\log \rho_A - \log \rho_B)] \geq 0$ is the quantum relative entropy~\cite{Vedral2002}. We notice that this expression is in agreement with previous results derived at the ensemble level for a single thermal reservoir~\cite{Esposito2010,Reeb2014} and for more general collisional dynamics~\cite{Strasberg2017,Cusumano2018,Rodrigues2019}{(for a recent review see Ref.~\cite{Ciccarello2021})}. Moreover, we notice that the total entropy changes in the environment may have extra terms as well due to other mechanisms implicit in the monitoring scheme and apart from the interaction with the system, as indeed happens in some decoherence models~\cite{Popovic2021}.

The entropy production during the trajectory $\gamma_{[0,\tau]}$ can then be defined as the sum of the changes in the entropy of the system plus the changes in entropy of the environment due to the entropy flow:
\begin{align} \label{eq:EP}
    S_\mathrm{tot}(\gamma_{[0,\tau]}) &= \Delta S(\gamma_{[0,\tau]}) + \sigma_\Lambda({\gamma}_{(0,\tau)}), \\  \label{eq:EP2}
    &= \Delta S(\gamma_{[0,\tau]}) - \sum_{r=1}^R \frac{Q^{(r)}_\Lambda(\gamma_{[0,\tau]})}{T_r}, 
\end{align}
where in the second line we split the entropy flow in contributions from each reservoir and used the identification of the stochastic heat in Eq.~\eqref{eq:heat}. Equations~\eqref{eq:EP} and \eqref{eq:EP2}  are the extension to quantum trajectories of the stochastic entropy production employed in stochastic thermodynamics~\cite{Seifert2012,VdB2015} and can be particularized for a number of cases of interest. For example in the isothermal situation ($R=1$), $1/T_r \equiv \beta$, using the first law in Eq.~\eqref{eq:work} we have:
\begin{equation} \label{eq:EP-isothermal}
S_\mathrm{tot}(\gamma_{[0,\tau]}) = \beta (W(\gamma_{[0,\tau]}) - \Delta F(\gamma_{[0,\tau]})),   
\end{equation}
where we introduced the stochastic non-equilibrium free energy changes along the trajectory, $\Delta F(\gamma_{[0,\tau]})) = \Delta E(\gamma_{[0,\tau]}) - T \Delta S(\gamma_{[0,\tau]})$. We recall that in the above equation the work done during the trajectory includes, in general, the three contributions highlighted in Eq.~\eqref{eq:work-split}.

It is worth mentioning at this point that one may be tempted to use Eq.~\eqref{eq:ef-probs} as a definition of the entropy flow to the environment $\sigma_\Lambda(\gamma_{(0,\tau)})$, without relying into Eq.~\eqref{eq:ef}. That would allow to e.g. consider more general diffusive trajectories beyond the $\sigma_\Lambda(\gamma_{(0,\tau)}) = 0$ cases. {This is an interesting route which merits further exploration. However,} the weakness of such an approach is that {it misses to constrain} the measurement operators in the backward process to be both normalized and able to reverse the trajectories{, so extra care would be needed}. This in turn may introduce extra irreversibility~\cite{Manikandan2019} and an extra dependence of $\sigma_\Lambda$ on the initial and final outcomes of the trajectories $n_0$ and $n_\tau$~~\cite{Manikandan2019b}.

Another interesting approach that bypass some of the problems regarding the definition of entropy production under diffusive measurements, involves an alternative definition of entropy production based on the Wigner function and Wigner entropy~\cite{Adesso2012}, the so-called Wigner entropy production rate~\cite{Santos2017}, which can be applied to general Gaussian setups. The Wigner entropy production coincides with the standard entropy production in the high temperature limit and {is well-suited for studying zero-temperature environments}. In Ref.~\cite{Belenchia2020} this approach was used to single out the entropy flux to the environment and the information rate gathered by the continuous monitoring process, hence extending the Sagawa-Ueda second-law inequality with information beyond discrete measurements for Gaussian systems, {which was} experimentally tested in an optomechanical setup~\cite{Rossi2020}.

\subsection{Fluctuation theorems}

The definition of the trajectory entropy change in Eq.~\eqref{eq:sys-entropy} and the microreversibility relation leading to Eq.~\eqref{eq:ef-probs}, imply that the entropy production becomes a measurement of irreversibility along trajectories~\cite{Horowitz2012,Horowitz2013,Manzano2015,Manzano2018b}, verifying the detailed fluctuation theorem:
\begin{equation} \label{eq:detailed}
    S_\mathrm{tot}(\gamma_{[0,\tau]}) = \log \left( \frac{P_\Lambda(\gamma_{[0,\tau]})}{P_{\tilde{\Lambda}}(\tilde{\gamma}_{[0,\tau]})} \right).
\end{equation}
Its average can be identified with the distinguishability between forward and time-reversed trajectories as in the classical case~\cite{Kawai2007}:
\begin{equation} \label{eq:KL}
    \langle S_\mathrm{tot}(\gamma_{[0,\tau]}) \rangle_\gamma = D_\mathrm{KL}[P_\Lambda(\gamma_{[0,\tau]}) || P_{\tilde{\Lambda}}(\tilde{\gamma}_{[0,\tau]})],
\end{equation}
where $D_\mathrm{KL}[\{p_n\} || \{q_n\} ] = \sum_n p_n \log(p_n/q_n)$ denotes the Kullback-Leibler divergence, which is non-negative $D_\mathrm{KL}[\{p_n\} || \{q_n\} ] \geq 0$ and becomes zero only for equal probability distributions. It is a classical version of the quantum relative entropy and measures the information lost when approximating $\{p_n\}$ with $\{q_n\}$~\cite{Cover2006}. 

From Eq.~\eqref{eq:detailed} it is immediate to obtain the integral fluctuation theorem:
\begin{eqnarray} \label{eq:integral}
\langle e^{-S_\mathrm{tot}(\gamma_{[0,\tau]})} \rangle_\gamma  = \sum_{\{ \gamma \}} P_{\tilde{\Lambda}}(\tilde{\gamma}_{[0,\tau]}) = 1,
\end{eqnarray}
which constrains several properties of the entropy production fluctuations, $P(S_\mathrm{tot})= \sum_{\{\gamma\}} P(\gamma_{[0,\tau]}) \delta(S_\mathrm{tot} - S_\mathrm{tot}(\gamma_{[0,\tau]}))$. For example Eq.~\eqref{eq:integral} implies an exponential tail for the probability of negative entropy events:
\begin{equation}
  P(S_\mathrm{tot} \leq -x) \leq e^{-x},   
\end{equation}
where $x\geq 0$. In the isothermal situation, Eq.~\eqref{eq:EP-isothermal}, this relation becomes a statement about the probability to extract work out of the system on the top of the free energy change, $P(W \leq \Delta F -x) \leq e^{-x}$, where again $x>0$, while Eq.~\eqref{eq:integral} becomes a generalized version of the Jarzynski equality~\cite{Jarzynski2011}. Moreover the fluctuation theorem in Eq.~\eqref{eq:integral} directly implies, by means of Jensen's inequality $ \langle e^X \rangle \geq e^{\langle X \rangle}$, the second-law inequality: 
\begin{equation}
\langle S_\mathrm{tot}(\gamma_{[0,\tau]}) \rangle_\gamma \geq \log \langle S_\mathrm{tot}(\gamma_{[0,\tau]}) \rangle_\gamma = 0.
\end{equation}

The detailed fluctuation theorem in Eq.~\eqref{eq:detailed} becomes a particularly stronger statement in the case of time-symmetric driving $\Lambda = \tilde{\Lambda}$, such as for constant Hamiltonian or some instances of periodic driving, leading to a non-equilibrium steady state of the system. In that cases, whenever the system starts in the long-time (asymptotic) steady state of the dynamics $p_{n_0} = \pi_{n_0}$, Eq.~\eqref{eq:detailed} reproduces the so-called Evans-Searles~\cite{Evans2002} or Gallavoti-Cohen~\cite{Gallavotti1995,Lebowitz1999} fluctuation theorem:
\begin{equation} \label{eq:ss-ft}
P(S_\mathrm{tot}) = P(-S_\mathrm{tot}) e^{S_\mathrm{tot}},
\end{equation}
which links the two tails of the probability density $P(S_\mathrm{tot})$ in the forward process. In the quantum scenario it has been typically used to describe the statistics of energy and other currents in nonequilibrium steady states~\cite{Esposito2009,Campisi2011}. In Ref.~\cite{Miller2021}, the fluctuation relation \eqref{eq:ss-ft} has been extended to transitions between thermal steady states for driving dynamics much slower than the system relaxation time-scales~\cite{Cavina2017}, and a similar identity for the joint probability of work and entropy production fluctuations, has been used to obtain finite-time corrections to the Carnot efficiency~\cite{Miller2020}.

It is worth remarking that the correct identification of the operators in the backward process through the micro-reversibility relation in Eq.~\eqref{eq:ef} is crucial to recover the correct expression of the stochastic entropy production and the integral fluctuation theorem in Eq.~\eqref{eq:integral}. Some attempts to define the stochastic entropy production omitting the TPM lead to the break of the integral fluctuation theorem by a efficacy-like parameter $\alpha$, such that $\langle e^{-S_\mathrm{tot}}\rangle = \alpha \leq 1$, which depends on the unravelling details~\cite{Leggio2013}. A similar situation arises for related definitions of an arrow of time indicator~\cite{Dressel2017,Harrington2019} where the efficacy term have been shown to be a form of so-called absolute irreversibility~\cite{Manikandan2019}. Such extra source of irreversibility can be avoided in the TPM framework presented here, while absolute irreversibility in general will not appear whenever the density operators for the initial states of forward and time-reversal processes share the same support~\cite{Funo2015}. Recently, other general fluctuation relations for quantum jump trajectories based on generic symmetries other than time-reversal, has been derived in the long-time limit using large deviation theory~\cite{Marcantoni2021}.

\section{Decomposition of entropy production}\label{dec_ep}

We have seen that the stochastic entropy production in Eq.~\eqref{eq:EP} contains two terms, one related to the change in entropy of the system, and another related to the entropy that is transferred to the environment. In the following we will see that the entropy production can be instead split in other meaningful ways, whose parts {separately} verify fluctuations theorems. Such splits refine our understanding of the second law at the level of fluctuations, since the contributions retain many of the nice properties and constraints mentioned before, and provide further insights on the energetics. We will focus in two particular splits of the entropy production: the first split is based on the adiabaticity of the evolution and comes from the context of stochastic thermodynamics for nonequilibrium transitions between steady states. The second split is instead a genuine partition of the entropy production along quantum trajectories based on the effects of the measurement and allows the characterization of quantum effects on the entropy production statistics.

\subsection{Adiabatic and Non-adiabatic Entropy Production}

The total entropy production can be decomposed in two components related to different sources of irreversibility arising in nonequilibrium transitions among steady states~\cite{Oono1998}. In the context of stochastic thermodynamics these contributions have been called the adiabatic and non-adiabactic entropy production~\cite{Esposito2010b,Esposito2010c}
\begin{equation} \label{eq:split}
    S_\mathrm{tot}(\gamma_{[0,\tau]}) = S_\mathrm{ad}(\gamma_{[0,\tau]}) + S_\mathrm{na}(\gamma_{[0,\tau]})
\end{equation}
and correspond, respectively, to the irreversibility arising from the breaking of detailed balance due to either nonequilibrium envrionmental conditions and external driving. They have been shown to verify separate fluctuation theorems~\cite{Esposito2010b}, generalizing a series of previous results~\cite{Hatano2001,Speck2005} regarding the heat needed to maintain a nonequilibrium steady state (house-keeping heat) and dissipated when driven far from it (excess heat), see e.g. the review~\cite{Seifert2012}. 

The split of the entropy production into adiabatic and non-adiabatic contributions along quantum jump trajectories [Eq.~\eqref{eq:split}] has been first considered in Ref.~\cite{Horowitz2013} and then extended to more general setups and generic quantum operations in Refs.~\cite{Manzano2015,Manzano2018b}. Here we will follow the general derivation reported there. We assume $\pi_\lambda$ to be a instantaneous invariant state of the Lindbladian, verifying  $\mathcal{L}_\lambda (\pi_\lambda)= 0$. Notice that although $\pi_\lambda$ would be unique in many cases of interest, we do not impose that condition generically. Moreover, it follows from the expressions for the Lindblad operators in the backward process in Eq.~\eqref{eq:backward-jump} that $\tilde{\pi}_{\lambda_{\tau -t}} = \Theta \pi_{\lambda_t} \Theta^\dagger$ is the instantaneous invariant state of the time-reversed dynamics, $\tilde{\mathcal{L}}_\lambda (\tilde{\pi}_\lambda) = 0$, where the generator $\tilde{\mathcal{L}}_\lambda$ is given in Eq.~\eqref{eq:backward-Lindblad}.

Following Refs.~\cite{Manzano2015,Manzano2018b} the split can be performed by introducing the nonequilibrium potential operator, namely $\Phi_\lambda \equiv - \log \pi_\lambda$ , which is a quantum version of the nonequilibrium potential used by Hatano and Sasa~\cite{Hatano2001}. The split can be performed if the jump operators verify the condition:
\begin{equation} \label{eq:noneq-condition}
[\Phi_\lambda, L_k(\lambda)] = \Delta \phi_k(\lambda) L_k(\lambda),
\end{equation}
for a set of real numbers $\{ \Delta \phi_k \}$, which also implies $[\Phi_\lambda, L_k^\dagger L_k] = 0$. The above equation implies that the Lindblad operators produce jumps (or coherent combinations thereof) in the $\pi_\lambda$ basis with a definite change in the nonequilibrium potential given by $\Delta \phi_k(\lambda)$. In addition, the Hamiltonian contribution in the Lindblad equation would need to verify $[H_\lambda, \phi_\lambda] = 0$, which imply that the jumps are in the energy $H_\lambda$ basis. However we remark that, in some cases, the later condition can be avoided by transforming the Lindblad equation to a rotating frame. 

When Eq.~\eqref{eq:noneq-condition} is verified, one can introduce the dual and dual-reversed processes, which are similar to the forward and backward processes respectively, but they are characterized by modified jump operators that verify relations similar to Eq.~\eqref{eq:backward-jump} but involving changes in the nonequilibrium potential (see appendix~\ref{app} for details). This allow us to obtain the probability of trajectories in the dual process, $P_\Lambda^{+}$, and the probability of time-reversed trajectories in the dual-reverse process $P_{\tilde{\Lambda}}^{+}(\tilde{\gamma}_{[0,\tau]})$. Then using similar methods than the ones used for obtaining the detailed fluctuation theorem in Eq.~\eqref{eq:detailed} one can verify the following relations that constitute the definition of the adiabatic and non-adiabatic entropy production along quantum trajectories:
\begin{align} \label{eq:detailed-ad}
 S_{\mathrm{ad}}(\gamma_{[0,\tau]}) &= \log \left( \frac{P_\Lambda(\gamma_{[0,\tau]})}{P_\Lambda^{+} (\gamma_{[0,\tau]})} \right) = \sigma_\Lambda(\gamma_{[0,\tau]}) + \Delta \Phi_\Lambda({R}_{[0,\tau]}), \\ \label{eq:detailed-na}
 S_{\mathrm{na}}(\gamma_{[0,\tau]}) &= \log \left(\frac{P_\Lambda(\gamma_{[0,\tau]})}{P_{\tilde{\Lambda}}^{+} (\tilde{\gamma}_{[0,\tau]})} \right) = \Delta S(\gamma_{[0,\tau]}) - \Delta \Phi_\Lambda({R}_{[0,\tau]}).   
\end{align}
Notice that while the non-adiabatic entropy production depends on the system entropy changes and hence on the initial and final trajectory outcomes $n_0$ and $n_\tau$, the adiabatic entropy production only depends on the environmental monitoring record $\gamma_{(0,\tau)}$. This is in accordance with the notion that the adiabatic entropy production is entirely due to environmental non-equilibrium constraints. 

The stochastic adiabatic entropy production $S_{\mathrm{ad}}(\gamma_{[0,\tau]})$ vanish {if the sole effect} of the entropy flow is to produce a modification in the state of the system, e.g. pushing it towards the instantaneous steady state $\pi_\lambda$. Then $\sigma_\Lambda(\gamma_{(0,\tau)}) = - \Delta \Phi(\gamma_{(0,\tau)})$, or alternatively $\Delta s_k = -\Delta \phi_k$, case in which all the entropy production is non-adiabatic~\cite{Manzano2018b}. {The most paradigmatic case in which this situation is verified is} in the presence of a single thermal reservoir. {However, as soon as various reservoirs with different temperatures or chemical potentials are considered, the adiabatic contribution becomes is in general non-zero, since some part of the entropy flow from one reservoir is damped into the others. In such general situations, some care must be taken to not confuse the total entropy production with the non-adiabatic part.} On the other hand, the stochastic non-adiabatic entropy production $S_{\mathrm{na}}(\gamma_{[0,\tau]})$ vanishes when the system is always maintained in the instantaneous steady-state of the dynamics {and the later is still far from equilibrium}. In such case all the entropy production is adiabatic, as situation that is verified e.g. in heat engines and refrigerators working continuously in non-equilibrium steady states~\cite{Kosloff2014}.

The above equations \eqref{eq:detailed-ad}-\eqref{eq:detailed-na} lead to relations analogous to Eq.~\eqref{eq:KL} for the average of the adiabatic and non-adiabatic entropy productions in terms of the Kullback-Leibler divergence: 
\begin{align}
\langle S_{\mathrm{ad}}(\gamma_{[0,\tau]}) \rangle_\gamma &= D_\mathrm{KL}[P_\Lambda(\gamma_{[0,\tau]}) || P_\Lambda^{+} (\gamma_{[0,\tau]})] \geq 0,    \\
\langle S_{\mathrm{na}}(\gamma_{[0,\tau]}) \rangle_\gamma &= D_\mathrm{KL}[P_\Lambda(\gamma_{[0,\tau]}) || P_{\tilde{\Lambda}}^{+} (\tilde{\gamma}_{[0,\tau]})] \geq 0,
\end{align}
and integral fluctuation theorems:
\begin{equation} \label{eq:integral-ad}
\langle e^{S_{\mathrm{ad}}(\gamma_{[0,\tau]})} \rangle_\gamma =1 ~~~,~~~ \langle e^{S_{\mathrm{na}}(\gamma_{[0,\tau]})} \rangle_\gamma = 1.    
\end{equation}
which generalize the Speck-Seifert~\cite{Speck2005} fluctuation theorem and the Hatano-Sasa relation~\cite{Hatano2001} respectively. Equations \eqref{eq:detailed-ad}-\eqref{eq:detailed-na}, together with \eqref{eq:integral-ad} constitute the quantum version of the fluctuation theorems for the adiabatic and non-adiabatic entropy productions obtained in Ref.~\cite{Esposito2010b}.

Some extra insight can be obtained by explicitly calculating the average non-equilibrium potential changes rate~\cite{Manzano2018b}:
\begin{eqnarray} \label{eq:av-phi}
\langle \Delta \dot{\Phi}_\Lambda(\gamma_{[0,\tau]}) \rangle_\gamma = \sum_k \langle L_k^\dagger L_k \rangle \Delta \phi_k = \tr[\Phi_\lambda \sum_k \mathds{D}_k(\rho_t)], ~~
\end{eqnarray}
where we used the commutation relation in Eq.~\eqref{eq:noneq-condition}. Using this expression we can now provide a closed expression for the average non-adiabatic entropy production rate as~\cite{Manzano2018b,Horowitz2014}:
\begin{equation}\label{eq:dot-sa}
 \langle \dot{S}_\mathrm{na}(t) \rangle_\gamma = \tr[\dot{\rho}_t (\log \pi_{\lambda_t} - \log \rho_t)],
\end{equation}
where we used the explicit form of the nonequilibrium potential operator $\Phi_{\lambda_t} = - \log \pi_{\lambda_t}$ and the fact that $[\Phi_{\lambda}, H(\lambda)] = 0$. {From Eq.~\eqref{eq:dot-sa}} it immediately follows that for quasi-static processes for which $\rho_t \simeq \pi_{\lambda_t}$ then $\langle \dot{S}_\mathrm{na} \rangle_\gamma \simeq 0$. Moreover, we observe that in the absence of driving, e.g. in relaxation processes, the above expression becomes the entropy production derived by Sponh, $\langle \dot{S}_\mathrm{na}(t) \rangle_\gamma = - \frac{d}{dt} S(\rho_t || \pi)$.

Using Eq.~\eqref{eq:av-heat} and the expression obtained before for the average entropy flow rate in Eq.~\eqref{eq:ef-average}, the average adiabatic entropy production rate turns out to be:
\begin{equation}
    \langle \dot{S}_\mathrm{ad}(t) \rangle_\gamma = \sum_{r} \sum_{k \in \mathcal{R}_r} \tr[\mathds{D}_k(\rho_t) \left( \sum_i \frac{\nu_i^{(i)}}{T_r} X_i(\lambda_t) - \Phi_{\lambda_t}\right)].
\end{equation}
It therefore follows that the adiabatic entropy production becomes zero if the nonequilibrium potential has the form $\Phi_{\lambda_t} = \sum_i \nu_r^i X_i^{(r)}(\lambda_t)$, {which is the case if the instantaneous steady-state is in equilibrium,} $\pi_\lambda \propto e^{-\sum_i \nu_i^{(r)} X_i/T_r}$. {This} is the case for a single reservoir, but cannot be verified whenever there are several reservoirs at different temperatures $T_r$ and e.g. chemical potentials $\mu_r$.

We remark that the conditions given in Eq.~\eqref{eq:noneq-condition} are stronger that the ones needed to ensure a well-defined time-reversed dynamics verifying the detailed fluctuation theorem in Eq.~\eqref{eq:detailed} for the total entropy production. Indeed, there are relevant cases where Eq.~\eqref{eq:detailed} is verified while condition \eqref{eq:noneq-condition} is not. An explicit example was analyzed in Ref.~\cite{Manzano2018b} for a periodically driven harmonic oscillator, where the coherent driving is weak enough to not modify the structure of the thermal dissipator, while inducing coherence in the steady state (see also the first example in Sec.~\ref{sec:examples} for a similar situation). Another relevant situation comprises cases with extended interacting systems and local dissipation. In that case the Lindblad operators associated to the local dissipators do not necessarily promote jumps in the basis of the global steady state of the system, due to the presence of a coherent coupling. Examples of this kind of dynamics are some models of quantum autonomous refrigerators~\cite{Linden2010,Brunner2014} and other thermal machines {used} for entanglement generation~\cite{Brask2015}. 

{The cumulants of the non-adiabatic entropy production along quantum jump trajectories has been explicitly obtained in Ref.~\cite{Miller2021} for slowly driven processes in contact with a single reservoir [where Eq.~\eqref{eq:noneq-condition} is typically satisfied] by constructing a two-variables moment generating function involving also the non-adiabatic work~\cite{Plastina2014}. These results were subsequently applied to quantum heat engines working within the slow-modulation regime to obtain bounds on the efficiency at finite time~\cite{Miller2021b}. }  

\subsection{Uncertainty and Martingale entropy production}

Very recently, a new decomposition of the entropy production has been proposed based on an extension of the so-called Martingale theory for entropy production~\cite{Chetrite2011,Neri2017,Neri2019,Chetrite2019,Neri2020} to quantum trajectories~\cite{Manzano2019,Manzano2021}. Although we will not discuss here Martingale theory for entropy production, we can fully introduce this decomposition of entropy production with the elements at hand. This decomposition results particularly useful in order to split the entropy production into classical and quantum contributions at the level of single trajectories, and highlights the entropic effects due to the end-point projections in the TPM scheme.

The decomposition of the entropy production is based on a different notion of the entropy of the system as given by the logarithm of the {quantum} fidelity between the density operator $\rho_\tau$ of the average dynamics and the stochastic wave function given by the SSE $\ket{\psi_\gamma(\tau)}$ just prior to the final application of the projectors $\{ \Pi_n^\tau \}$. That is~\cite{Manzano2019,Manzano2021}:
\begin{eqnarray} \label{eq:Uhlmann-entropy}
S_\psi(\tau) \equiv  -\log \langle \psi_\gamma(\tau) | \rho_\tau | \psi_\gamma(\tau) \rangle.
\end{eqnarray}
Here we assumed for simplicity that the stochastic evolution has no extra classical sources of uncertainty and is hence described by the SSE. However the definition can be extended to the case of the SME as $S_\rho(\tau) \equiv - \log \tr[\rho_\tau \rho_\gamma(\tau)]$, which is no longer the logarithm of a precise fidelity in general {(like Uhlmann's fidelity for mixed states)}. In any case, we notice that this notion of entropy differs from the system entropy used in Eq.~\eqref{eq:sys-entropy}. In particular Eq.~\eqref{eq:Uhlmann-entropy} reduces to the standard surprisal only when the stochastic wave function at time $\tau$ remains in a eigenstate of the density operator at the final time $\rho_\tau$, i.e. when $\langle \psi_\gamma(\tau)| \Pi_k^{\tau} |\psi_\gamma (\tau) \rangle = \delta_{l,k}$ for some $l$, which is not the case in general. As a consequence, by averaging Eq.~\eqref{eq:Uhlmann-entropy} along trajectories we do not recover von Neumann entropy $\langle S_\psi(\tau) \rangle \neq S_\mathrm{vN}(\rho_\tau)$.

Using the notion of entropy above we can then decompose the entropy production as:
\begin{equation} \label{eq:ep-split2}
S_\mathrm{tot}(\gamma_{[0,\tau]}) = S_\mathrm{unc}(\gamma_{[0,\tau]}) + S_\mathrm{mar}(\gamma_{[0,\tau]}).
\end{equation}
The first term is called the uncertainty entropy production and reads:
\begin{eqnarray} \label{eq:uncerainty}
S_\mathrm{unc}(\gamma_{[0,\tau]}) = -\log(p_{n_\tau}^\tau) - S_\psi(\tau),
\end{eqnarray}
where {$S_\psi$} is the fidelity-based entropy introduced in Eq.~\eqref{eq:Uhlmann-entropy}. Hence $S_\mathrm{unc}(\gamma_{[0,\tau]})$ corresponds to the part of the entropy production due to the final projection of the monitored system. The uncertainty entropy production is of quantum origin and becomes zero in the classical case, since in that case the system state is always an eigenstate of its density operator (and of any other observable). It can be interpreted as a disturbance due to the final projective measurement when the state $\ket{\psi_\gamma(\tau)}$ shows an intrinsic (quantum) uncertainty. It is worth remarking in this context that $S_\mathrm{unc}(\gamma_{[0,\tau]})$ is non-zero even when the projectors $\{\Pi_{k}^\tau \}$ are chosen to be in the basis of  $\rho_\tau$. 

In addition $S_\mathrm{unc}$ can be shown {to be} bounded by the log-ratio of the minimum and maximum eigenvalues of the density operator $\rho_\tau$~\cite{Manzano2019}:
\begin{eqnarray}
\log \left( \frac{p_{\min}^\tau}{p_{\max}^\tau} \right) \leq S_\mathrm{unc}(\gamma_{[0,\tau]}) \leq \log \left( \frac{p_{\max}^\tau}{p_{\min}^\tau} \right),
\end{eqnarray}
where we denoted $p_{\max}^\tau = \max_k p_k^\tau$ and $p_{\min}^\tau = \min_k p_k^\tau$.

The second term in Eq.~\eqref{eq:ep-split2} has been called the martingale entropy production, because it is an exponential martingale, a particularly strong property which, in particular, also implies the integral fluctuation theorem. The explicit form of the martingale entropy production reads:  
\begin{equation} \label{eq:martingale-ep}
 S_\mathrm{mar}(\gamma_{[0,\tau]}) = S_\psi(t) + \log p_{n_0} + \sigma_\Lambda(\gamma_{(0,\tau)}),    
\end{equation}
which remarkably does not depend on the end-point of the trajectory, i.e. the outcome of the final projective measurement $n_\tau$. The entropy production in Eq.~\eqref{eq:martingale-ep} represents somehow a ``classicalized" or ``smoothed" version of the entropy production $\Delta S_\mathrm{tot}(\gamma_{[0,\tau]})$ on which (at least part of) the quantumness is not present anymore, in the sense that the part due to the intrinsic uncertainty in the system state has been removed. As can be appreciated the martingale entropy production $S_\mathrm{mar}(\gamma_{[0,\tau]})$ is an extensive quantity, in contrast to $S_\mathrm{unc}(\gamma_{[0,\tau]})$, since it contains the entropy flow $\sigma_\Lambda(\gamma_{(0,\tau)})$.  
In Refs.~\cite{Manzano2020,Manzano2021} both the uncertainty and the martingale entropy productions have been shown to verify a integral fluctuation theorem in both steady-state and generic evolutions:
\begin{equation}
    \langle e^{-S_\mathrm{unc}(\gamma_{[0,\tau]})} \rangle_\gamma = 1 ~~,~~  \langle e^{-S_\mathrm{mar}(\gamma_{[0,\tau]})} \rangle_\gamma = 1,
\end{equation}
which in particular imply the second-law-like inequalities:
\begin{equation} \label{eq:inequalities}
    \langle S_\mathrm{unc} (\gamma_{[0,\tau]}) \rangle_\gamma \geq 0 ~~,~~ \langle S_\mathrm{mar} (\gamma_{[0,\tau]}) \rangle_\gamma \geq 0,
\end{equation}
together with the other properties of the integral fluctuation theorems regarding the negative tails of the distributions of both $S_\mathrm{unc}(\gamma_{[0,\tau]}$ and $S_\mathrm{mar}(\gamma_{[0,\tau]})$.
The non-trivial fact that $\langle S_\mathrm{unc}(\gamma_{[0,\tau]}) \rangle \geq 0$ also implies that the total entropy production can be lower bounded by the martingale one:
\begin{equation} \label{eq:bound}
   \langle S_\mathrm{tot} (\gamma_{[0,\tau]}) \rangle_\gamma \geq \langle S_\mathrm{mar} (\gamma_{[0,\tau]}) \rangle_\gamma.
\end{equation}
The above inequality represents an useful bound since $S_\mathrm{mar}(\gamma_{[0,\tau]})$ does not depend on the final projection $\Pi_{n_\tau}^\tau$ and might hence be obtained in real time only from the monitored record $\gamma_{(0,\tau)}$ and the initial preparation of the system state. Indeed $S_\mathrm{mar}(\gamma_{[0,\tau]})$ becomes particularly crucial when considering stopping times (e.g. first-passage times, escape times, etc) or gambling strategies, since in those cases one would like to decide to stop (or not) the process before introducing any disturbance into the system~\cite{Manzano2021}. We remark that the bound in~\eqref{eq:bound} becomes tight in the long time limit, since $S_\mathrm{unc}(\gamma_{[0,\tau]})$ is bounded, while $S_\mathrm{mar}(\gamma_{[0,\tau]})$ is extensive in time and hence $\langle S_\mathrm{tot}(\gamma_{[0,\tau]}) \rangle_\gamma \simeq \langle S_\mathrm{mar}(\gamma_{[0,\tau]}) \rangle_\gamma$ when $\tau \rightarrow \infty$.

\section{Simple Examples} \label{sec:examples}

\subsection{Driven two-level system in a thermal environment}

As a first example we consider a single two-level system driven by a coherent field in contact with a thermal environment at a finite temperature $T$, whose emission and absorption of excitations (e.g. photons) are monitored. We denote the Hamiltonian of the two-level system (without driving) as $H_S= \omega \ket{1}\bra{1}$, with system computational basis $\{\ket{0}, \ket{1}\}$. The driving is assumed to be weak and resonant with the two-level system. It induces an extra time-dependent term reading $V(t)=\epsilon (e^{-i\omega t} \sigma_+ + e^{i\omega t} \sigma_- )$, with $\sigma_{-} \equiv \ket{0}\bra{1}$ and $\sigma_{+} \equiv \ket{1}\bra{0} = \sigma_{-}^\dagger$ and driving strength $\epsilon \ll \omega$. In order to make contact with the thermodynamic framework for trajectories, we identify the control parameter as $\lambda (t) = \epsilon e^{i \omega t}$, and hence $\Lambda$ represents a cyclic protocol running during an arbitrary interval of time $[0,\tau]$. Moreover, we assume that the driving is instantaneously switched on at the beginning of the protocol and switched off at time $\tau$, that is $\lambda(0) = \lambda(\tau) = 0$.

The master equation governing the unconditional dynamical evolution of the dissipative-driven system is given by Eq.~\eqref{eq:Lindblad}, with $H(\lambda) = H_S + V(\lambda)$ and two Lindblad (jump) operators, ($k = +,-$), associated to  emission and absorption events reading: 
\begin{equation}\label{eq:lindblad1}
L_{-} = \sqrt{\Gamma_0 (\bar{n}+1)}~ \sigma_- ~~~;~~~    L_{+} = \sqrt{\Gamma_0~\bar{n}}~ \sigma_+
\end{equation}
which verify $L_{-} = e^{-\omega/T} L_{+}^\dagger$, with $\Gamma_0$ the spontaneous emission rate and $\bar{n} = (e^{\omega/T} + 1)^{-1}$ the average number of excitations in the thermal (bosonic) environment with frequency $\omega$. The above master equation in Lindblad form can be derived using standard methods in open quantum systems such as the Born-Markov and Secular approximations, and possess a single nonequilibrium steady state $\pi(t)$ which follows a closed unitary orbit, $\dot{\pi} = -i[H_S, \pi]$, due to the presence of coherences in the $H_S$ basis~\cite{Breuer2003}.

For assessing the thermodynamics during trajectories within the TPM, we assume the initial state of the system to be sampled from the local Gibbs state of the system before the driving is applied, $\rho_0 = e^{-\beta H_S}/Z$ with $Z= \tr[e^{-\beta H_S}] = 1 + e^{-\beta \omega}$, at the (inverse) environmental temperature $\beta = 1/T$. Therefore the two possible initial states of the system are either $\ket{0}$ or $\ket{1}$ with probabilities $p_0 = 1/Z$ and $p_1 = e^{-\beta \omega}/Z$, {that is, $\Pi_n^0 = \{ \ket{0}\bra{0}, \ket{1}\bra{1}$ for $n= 0,1$}.
As for the final projectors, we consider the simplest situation where they are given by the basis of the density operator at the final time, $\mathcal{E}(\rho_0)$, i.e. no extra disturbance at the unconditional level {($[\Pi_m^\tau, \mathcal{E}(\rho_0)]=0$)}.

\begin{figure*}[tb]
    \centering
    \includegraphics[width=1.0 \linewidth]{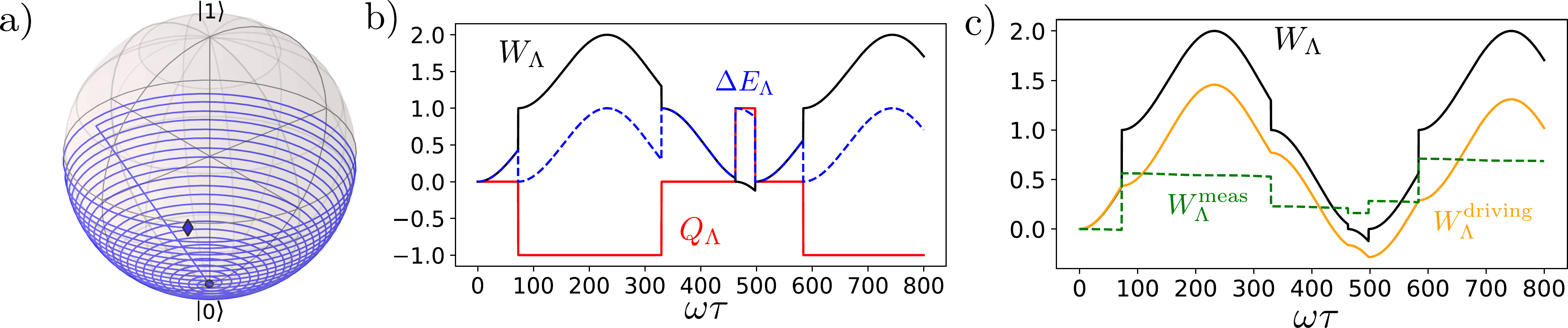}
    \caption{a) Sample trajectory evolution of the stochastic wave function in the surface of the Bloch sphere, starting in $\ket{0}$ (blue dot) and reaching a coherent state near the equator (blue diamond) at time {$t \sim \Gamma_0^{-1}$} before the final projection. During the trajectory a single emission is recorded, leading to a jump to the ground state (straight line). b) Energy change (dashed blue), work (solid black) and heat (solid red), {and} c) total work (solid black), driving work (solid orange) and measurement work (dashed green) along a single trajectory of the two-level system evolution as a function of the final time $\tau$ (including the final measurement). In all plots energetic quantities are given in $\hbar \omega$ units. Here  $\Gamma_0 = 0.001 \omega$, and $\epsilon = 0.01 \hbar \omega$, $k_B T = 5 \hbar \omega$.
    }
    \label{fig:qubit}
\end{figure*}

When the emission and absorption processes are (efficiently) monitored during the evolution, the state of the system conditioned to detections can be described by the stochastic Schr\"odinger equation:
\begin{align}
d{\ket{\psi_\gamma(t)}} = & dt\left[-i H[\lambda(t)] +\frac{\Gamma_0}{2} (\langle \sigma_+ \sigma_- \rangle - \sigma_+ \sigma_-)\right]\ket{\psi_\gamma(t)} \nonumber \\ 
&+ dN_{-} \left( \frac{\sigma_-}{\sqrt{\langle \sigma_+ \sigma_- \rangle}} - \mathds{1} \right) \ket{\psi_\gamma(t)} \nonumber \\ 
&+ dN_{+} \left( \frac{\sigma_+}{\sqrt{\langle \sigma_- \sigma_+ \rangle}} - \mathds{1} \right) \ket{\psi_\gamma(t)}.
\end{align}
where the Poissonian stochastic increments $dN_{\pm}(t)$ are zero almost all the time interval except when {any} jump of type $k=\pm$ is detected, in which case they become 1. An example of the evolution of the state of the system in the Bloch sphere of the two-level system is provided in Fig.~\ref{fig:qubit}a. {A} trajectory starts in the blue point marking the ground state, and is driving up along the surface of the sphere while rotating at frequency $\omega$. At some point an emission is detected producing a jump in the system back to the ground state (straight line), followed by a second period of rotation where the system is again directed towards the equator of the sphere up to the diamond point. 

We also notice that in some situations (e.g. photocounting) detection of absorption events $dN_{+}$ may be difficult to implement and hence require some detection engineering, like e.g. using an auxiliary highly unstable third level $\ket{2}$ from which emissions $\ket{2} \rightarrow \ket{1}$ can be photocounted~\cite{Elouard2017b}. In superconducting qubits this problem can be overcome by implementing a high-precision detection of the temperature variations due to the jumps that are produced in a resonator acting as the thermal reservoir~\cite{Karimi2020}. 

The jump operators \eqref{eq:lindblad1} do not depend on the control parameter $\lambda$, and promote jumps in the bare Hamiltonian basis $H_S$, i.e. 
$[H_S, L_{\pm}] = \pm \omega L_{\pm}$, in accordance to Eq.~\eqref{eq:energy-jumps} for energy jumps. It can be easily checked that the dissipative part of the {unconditional} evolution, {$\mathds{D}_+(\rho) + \mathds{D}_-(\rho)$}, has a single invariant state of Gibbs form, $e^{-H_S/ T}/Z$ (but the steady state of the {entire} dissipative-driven dynamics is not of Gibbs form). Therefore we can identify the energy change in the system during the jumps as $\Delta E_{\pm} = \pm \omega$, {corresponding} to a photon absorbed (emitted) from (into) the environment. In addition we observe that since the set of Lindblad operators is complete in the sense introduced in Sec.~\ref{sec:jumps}, we obtain the local detailed balance for the jump operators in Eq.~\eqref{eq:ldb} with associated entropy changes:
\begin{equation}
    \Delta s_{\pm} = \pm \frac{\omega}{T} = - \frac{\Delta E_{\pm}}{T}, 
\end{equation}
confirming that the energy quanta exchanged with the reservoir during the jumps can be interpreted as heat. This is all what we need to characterize the heat over trajectories. Substituting the form of $ \Delta s_{\pm}$ in Eqs.~\eqref{eq:ef-jumps} and \eqref{eq:heat}, we obtain:
\begin{equation}
Q_{\Lambda}(\gamma_{[0,\tau]}) = \int_0^\tau \omega (dN_{+} - dN_{-}) = \omega [N_+(\tau) - N_{-}(\tau)],
\end{equation}
thus the heat along the trajectory is the net energy absorbed from the reservoir during the jumps, i.e. energy absorbed with jumps of type $k=+$ minus energy released in the jumps of type $k=-$, during the interval $[0,\tau]$.

We now turn our attention to the total work, $W_\Lambda(\gamma_{[0,\tau]}) = \Delta E_\Lambda(\gamma_{[0,\tau]}) - Q_{\Lambda}(\gamma_{[0,\tau]})$ and its three contributions as introduced in Eq.~\eqref{eq:work-split}. First we notice that since there is a single conserved quantity (the energy) the chemical work contribution associated to extra conservation laws is zero in this case. The driving contribution can be calculated from Eq.~\eqref{eq:drive-work} and reads:
\begin{align}
W_\Lambda^{\mathrm{driving}}(\gamma_{[0,\tau]}) = & \tr[\Pi_n^0 V(0)] - \tr[\rho_\gamma(\tau) V(\tau)] \\ 
&-i\omega \epsilon \int_0^ \tau dt (e^{-i\omega t} \langle \sigma_+ \rangle - e^{i\omega t} \langle \sigma_- \rangle), \nonumber
\end{align}
where {the two terms in the first line} correspond respectively to the work needed to switch on and off the driving at the beginning and end of the protocol. Since the initial state is diagonal in the $H_S$ basis, we have $\tr[ \Pi_n^0 V(0)] = 0$, i.e. no work is needed to switch on the driving. Moreover since the steady state also verifies $\tr[V \pi] = 0$, we have also zero switch- off work cost in this case.

The work contribution due to continuous measurement is, from Eq.~\eqref{eq:meas-work}:
\begin{align}
  &W_\Lambda^{\mathrm{meas}}(\gamma_{[0,\tau]}) = \tr[H(\lambda_\tau) \Pi_m^{\tau}] - \tr[H(\lambda_\tau) \rho_\gamma(\tau)] \nonumber \\ &- \int_0^\tau dt \Gamma_0 \left( \frac{\langle \{H, \sigma_+ \sigma_- \} \rangle}{2} - E_\gamma \langle \sigma_+\sigma_- \rangle \right)  \nonumber \\ 
&+ \int_0^\tau dN_- \Gamma_0 (\bar{n} + 1) \left(\frac{1}{2} \frac{\langle \{H, \sigma_+ \sigma_- \} \rangle}{\langle \sigma_+ \sigma_- \rangle} - E_\gamma \right) \nonumber \\
&+ \int_0^\tau dN_+ \Gamma_0 \bar{n} \left(\frac{1}{2} \frac{\langle \{H, \sigma_- \sigma_+ \} \rangle}{ \langle \sigma_- \sigma_+ \rangle} - E_\gamma \right), 
\end{align}
where we recall that {the first line is due to the final projective measuremnt at time $\tau$, and} $E_\gamma(t) = \langle H(\lambda_t) \rangle$ is the instantaneous expected energy of the driven two-level system conditioned on the continuous measurement record.

In Fig.~\ref{fig:qubit}b we show the energy change $\Delta E_\Lambda(\gamma_{[0,\tau]})$, heat $Q_\Lambda(\gamma_{[0,\tau]})$ and total work $W_\Lambda(\gamma_{[0,\tau]})$ evaluated along a sample trajectory starting in the ground state (the first part of the trajectory is depicted in the Bloch sphere in Fig.~\ref{fig:qubit}a. The energy changes in the qubit reveal the Rabi oscillations followed by the two-level system during the no-jump periods, which are interrupted by the jumps associated to the emission and adsorption events. Heat is exchanged with the thermal environment only during the jumps, and we can appreciate three emission events (downstairs jumps) and two absorption ones (upstairs jumps). Instead, work is realized or extracted during both no-jump periods and jumps, contrary to the classical case. In Fig.~\ref{fig:qubit}c the total work is split into the driving $W_\Lambda^\mathrm{driving}(\gamma_{[0,\tau]})$ and measurement $W_\Lambda^\mathrm{meas}(\gamma_{[0,\tau]})$ contributions introduced in Sec.~\ref{subsec:work}, which are shown for the same trajectory. The driving contribution is continuous during the whole trajectory and shows a smooth behavior between jumps, reflecting the work performed to generate the Rabi oscillations. The jumps instead do not contribute to the driving work, but its detection produce cusps on it. On the other hand, the measurement work suffers abrupt changes owing to its stochastic nature and it is monotonous and slightly decreasing between jumps. This reflects the fact that when no jumps are detected the system is more likely in its ground state implying that less work have been employed in driving it from its initial (ground) state. In a similar way, detection of a emitted excitation suddenly increases the conditional work, since we just learn that energy (provided previously by the drive as work) has been dissipated into the environment, and the other way around for absorption events. 

\begin{figure}[tbh]
    \centering
    \includegraphics[width=0.8 \linewidth]{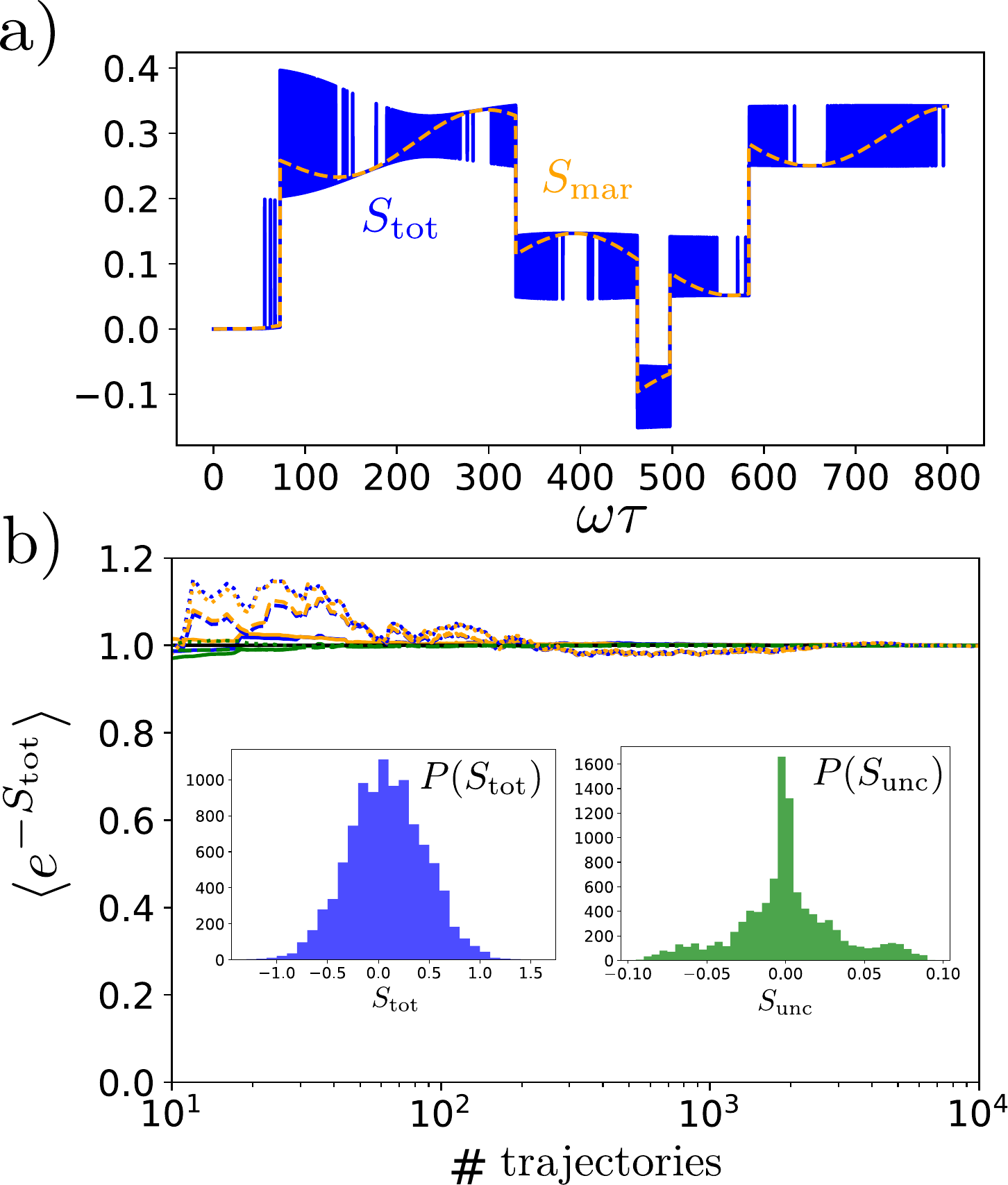}
    \caption{a) Total entropy production (blue) and martingale entropy production (dashed orange), along a single trajectory of the two-level system evolution as a function of the final time $\tau$ (including the final measurement). b) Convergence of integral fluctuation theorems $\langle e^{-S_\mathrm{tot}} \rangle_\gamma$ (blue), $\langle e^{-S_\mathrm{mar}} \rangle_\gamma$ (orange) and $\langle e^{-S_\mathrm{unc}} \rangle_\gamma$ (green), as a function of the number of trajectories employed in the simulations for different times $\omega \tau = 100$ (solid) $\omega \tau = 500$ (dashed) and $\omega \tau = 800$ (dotted). Inset: Probability distributions for the total (left), and uncertainty entropy production (right) evaluated at final time $\tau$. In all plots energetic quantities are given in $\hbar \omega$ units. We used again $\Gamma_0 = 0.001 \omega$, $\epsilon = 0.01 \hbar \omega$, $k_B T = 5 \hbar \omega$ and $10^4$ trajectories for the simulations.
    }
    \label{fig:qubit-ft}
\end{figure}

We are also interested in the stochastic entropy production in this example and its fluctuations and {contributions}. The stochastic entropy production, as given by Eq.~\eqref{eq:EP}, reduces to {the} expression for a single thermal reservoir $S_\mathrm{tot}(\gamma_{[0,\tau]}) = \beta (W(\gamma_{[0,\tau]} - \Delta F)$ given in Eq.~\eqref{eq:EP-isothermal}, where $\Delta F(\gamma_{[0,\tau]}) = \Delta E(\gamma_{[0,\tau]}) -  T \Delta S(\gamma_{[0,\tau]})$ is the stochastic non-equilibrium free energy. We plot $S_\mathrm{tot}(\gamma_{[0,\tau]})$ in blue in Fig.~\ref{fig:qubit-ft}a by assuming virtual end measurements performed at each value of the final time $\tau$. High-frequency noisy changes in the entropy production reflect the quantum fluctuations associated to the final measurement and pinpoints the intervals in which the stochastic wave function $\ket{\psi_\gamma(t)}$ is in a superposition of the density operator eigenstates. The decomposition into uncertainty and martingale components in Eq.~\eqref{eq:ep-split2} is shown by means of the dashed orange line. As can be appreciated, the above quantum fluctuations are absent in the martingale entropy production, leading to a smooth version of the total entropy production, showing abrupt changes only corresponding to the energy jumps in the two-level system. The effect of the Rabi oscillations can be also appreciated in the evolution of $S_\mathrm{mar}(\gamma_{[0,\tau]})$ during the no-jump intervals. The local maxima and minima during such intervals correspond to zero quantum fluctuations, i.e. the stochastic wave function $\ket{\psi_\gamma(\tau)}$ prior to measurement becomes one of the eigenstates of the density matrix $\rho(\tau)$ at that time. The uncertainty entropy production (not shown in the figure) is the difference between $S_\mathrm{tot}(\gamma_{[0,\tau]})$ and $S_\mathrm{mar}(\gamma_{[0,\tau]})$, hence capturing only the high-frequency quantum fluctuations produced in the final measurement. In Fig.~\ref{fig:qubit-ft}b we show the convergence of the three different fluctuation theorems for the total $\langle e^{-S_\mathrm{tot}} \rangle_\gamma$ (blue), martingale $\langle e^{-S_\mathrm{mar}} \rangle_\gamma$ (orange) and uncertainty $\langle e^{-S_\mathrm{unc}} \rangle_\gamma$ (green) entropy production as a function of the number of trajectories employed in the simulations for three different final times $\omega \tau$ (see caption). We can see a good convergence in the three cases, being the uncertainty entropy production fluctuation theorem the one that converges more quickly to $1$. In the inset we provide estimations of the total entropy production and uncertainty probability distributions, $P(\Delta S_\mathrm{tot})$ and $P(S_\mathrm{unc})$ respectively, for a fixed final time $\omega \tau = 2500$. As we can appreciate the entropy production distribution is close to a Gaussian with positive mean $\langle \Delta S_\mathrm{tot} \rangle_\gamma \geq 0$. The uncertainty entropy production instead displays a large peak but also some secondary peaks at both positive and negative sides. Since $S_\mathrm{unc}$ is a bounded non-extensive quantity over time, the whole distribution is closer to zero as compared with $P(\Delta S_\mathrm{tot})$.

On the other hand, the split into adiabatic and non-adiabatic entropy production contributions in Eq.~\eqref{eq:split} cannot be implemented in this setup. This is because the condition in Eq.~\eqref{eq:noneq-condition} is not verified, since the jumps promoted by the Lindblad operators are not eigenoperators of the steady state $\pi(\lambda)$ and hence of the nonequilibrium potential $\Phi(\lambda)$, breaking the fluctuation theorems in Eqs.~\eqref{eq:integral-ad}. This sets apart this configuration from any classical jump process, for which the adiabatic and non-adiabatic entropy production are always well defined and remarks the importance of quantum effects in the fluctuations. In this respect, the situation is very similar to the weakly driven cavity mode dissipating into a thermal environment considered in Ref.~\cite{Manzano2018b}, where the break of the split leads to an adiabatic entropy production which would be negative on average.

\subsection{Non-demolition monitoring of a driven superconducting qubit}

As a second example we consider the case of a coherently driven superconducting qubit over which a dispersive measurement of energy is applied~\cite{Alonso2016,DiStefano2018}. This setup has been experimentally implemented in Ref.~\cite{Naghiloo2018} using a transmon qubit coupled to a 3D aluminium cavity to implement a Maxwell demon. No further contact with a thermal reservoir is assumed. The qubit does not exchange energy with the cavity mode, so that it acts as a dephasing environment for the qubit, induced by the continuous (homodyne) measurement of the cavity.

The Hamiltonian of the qubit is again $H(\lambda) = H_S + V(\lambda)$, where the free system Hamiltonian reads $H_S = -\frac{\omega_q}{2} \sigma_z$, and the driving contribution $V(\lambda) = -i \sigma_y \lambda$ with $\lambda(t) = \Omega_R \cos(\omega_q t)$ the control parameter inducing the driving. We assume again a weak driving $\Omega_R \ll \omega_q$. The coupling with the cavity mode is assumed to be of the form $H_\mathrm{int} = -\xi \sigma_z a^\dagger a$, with $a$ and $a^\dagger$ the ladder operators of the cavity mode, so that $[H_S, H_\mathrm{int}] = 0$. The cavity has energy $H_c = \omega_c a^\dagger a$ and is coherently probed to acquire information about the state of the system~\cite{Naghiloo2018}. {This leads to a diffusive trajectory over the qubit, with a noisy output signal proportional to $\sigma_z$.}

In this situation the unconditional state evolution of the system can be described by a Lindblad equation implementing pure dephasing over the qubit, with a single Lindblad operator $L_z = \sqrt{\kappa} \sigma_z$, where $\kappa$ is a constant characterizing the strength of the measurement. We note that since we have a single self-adjoint Lindblad operator, $L_z = L_z^\dagger$, the micro-reversibility relation for diffusive trajectories stated in Sec.~\ref{subsec:micro-diffusive} holds. Moreover there {the} steady state of the evolution {is not unique} in this case: any state of the qubit diagonal in the $H_S$ basis is invariant under the action of the environment. In particular the maximally mixed state $\pi = \mathds{1}/2$ is also an invariant state, that is, the map generating the dynamics is unital.

\begin{figure}[tbh]
    \centering
    \includegraphics[width=0.7 \linewidth]{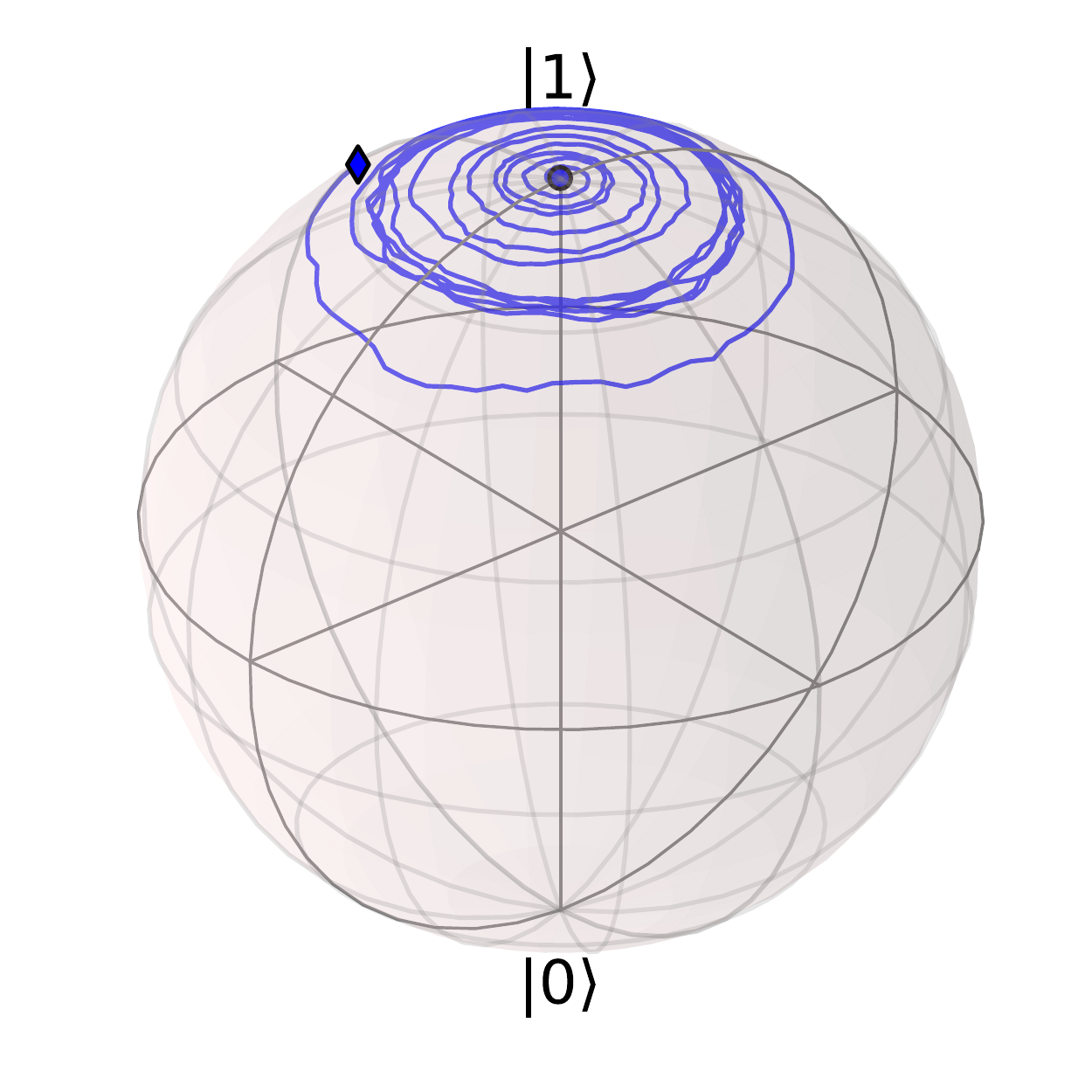}
    \caption{Initial moments for a sample trajectory evolution of the stochastic wave function in the surface of the Bloch sphere for the diffusive case. The qubit starts in the excited state $\ket{1}$ (blue dot) and drifts downward while rotating along the $z$ axis, leading to a coherent state at time $t= \omega_q^{-1}$ (blue diamond) before the final projection is applied. Parameters of the simulation: $\kappa = 0.001 \omega$, $\Omega_R = 0.01 \omega_q$.
    }
    \label{fig:dif-traj}
\end{figure}

A single current proportional to the qubit energy {corresponds to the measurement record, namely}  $\gamma_{(0,\tau)}=\{I(t); 0 \leq t \leq \tau\}$, with $I(t) = 2 \kappa \langle \sigma_z \rangle dt + dw(t)$ and the Wiener increment satisfies the rules  $\langle dw(t) dw(t^\prime)\rangle_\gamma = \delta(t-t^\prime) dt$ and $\langle dw(t) \rangle_\gamma = 0$. Taking into account the measurement record, the following stochastic master equation for the evolution of the qubit conditioned on the continuous measurement is obtained:
\begin{align} \label{eq:sme-qubit}
d \rho_\gamma = &-i[H, \rho_\gamma] dt - \frac{1}{2}[L_z,[L_z,\rho_\gamma] dt  \nonumber \\
&+ (L_z \rho_\gamma + \rho_\gamma L_z - 2 \langle L_z \rangle \rho_\gamma) dw,
\end{align}
where we assume efficient detection of the qubit by reading-out the cavity. 
Analogously to the previous example, we consider the initial state of the system at thermal equilibrium, $\rho_0 = e^{-\beta H_S}/Z$, from which we sample the initial state of the trajectories (in the $H_S$ basis), according to the probabilities $p_0 = e^{-\beta \omega_q/2}/Z$ and $p_1 = e^{-\beta \omega_q/2}/Z$ with $Z= e^{-\beta \omega_q/2} + e^{\beta \omega_q/2}$. We also consider that the final measurement of the TPM is performed in the $\mathcal{E}(\rho_0)$ basis. In the long time limit the state of the system approaches the fully mixed state $\mathds{1}/2$. A sample trajectory generated by the stochastic master equation in Eq.~\eqref{eq:sme-qubit} is shown in Fig.~\ref{fig:dif-traj}, to be compared with Fig.~\ref{fig:qubit}a. We note that since the initial state of the system is pure, and efficient detectors are considered, the evolution remains in the surface of the Bloch sphere also here (non-efficient detection would lead to excursion inside the sphere volume).

\begin{figure}[tb]
    \centering
    \includegraphics[width=0.8 \linewidth]{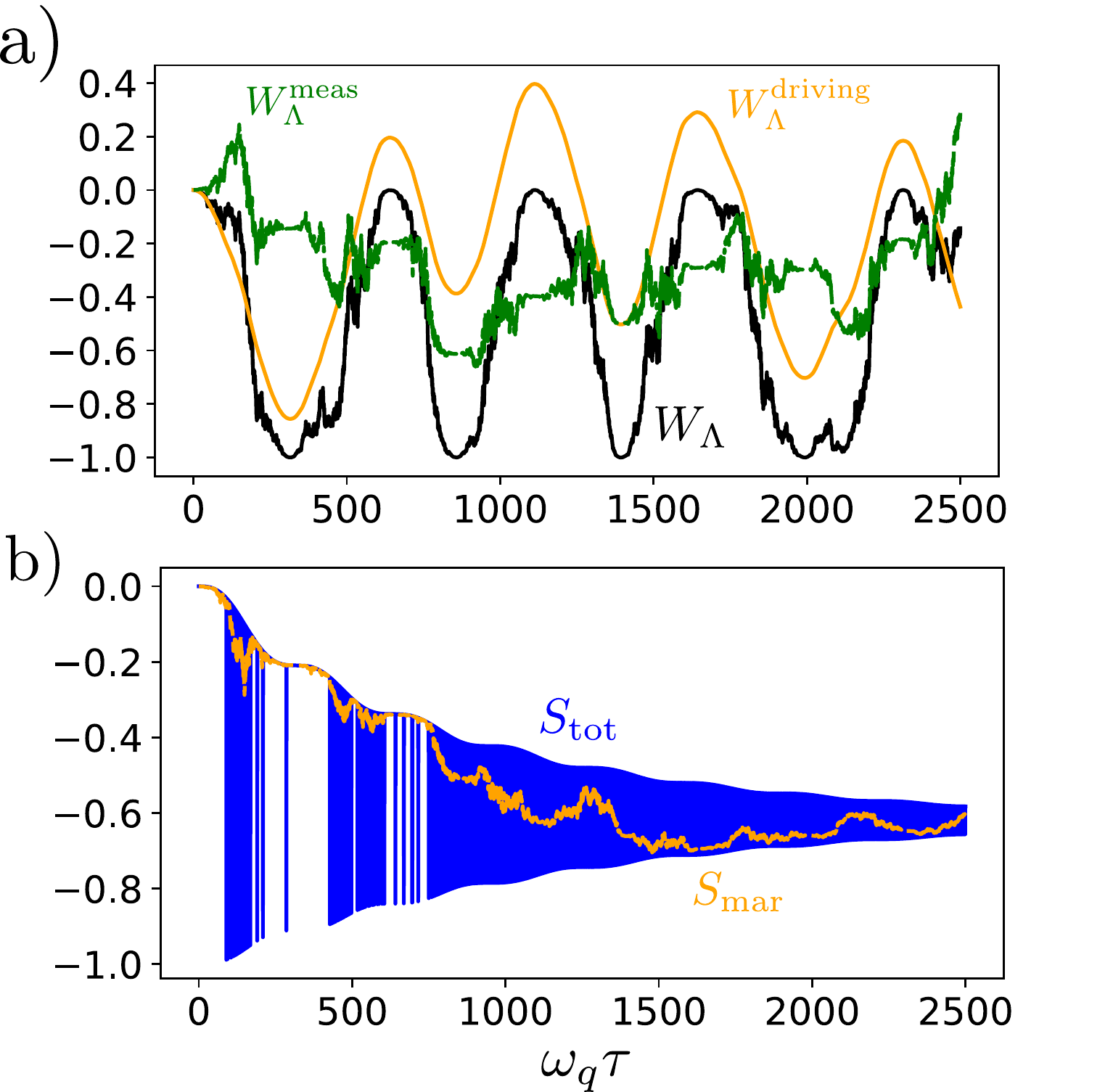}
    \caption{a) Total (stochastic) work $W_\Lambda(\gamma_{[0,\tau]})$ and contributions from driving $W_\Lambda^\mathrm{driving}(\gamma_{[0,\tau]})$ and measurement $W_\Lambda^\mathrm{meas}(\gamma_{[0,\tau]})$ as a function of the final time $\tau$, during a sample trajectory starting in the excited state. b) Total entropy production $S_\mathrm{tot}(\gamma_{[0,\tau]})$ and martingale entropy production $S_\mathrm{mar}(\gamma_{[0,\tau]})$ during the same trajectory.
    Parameters of the simulation: $\kappa = 0.001 \omega$, $\Omega_R = 0.01 \omega_q$.
    }
    \label{fig:work-diffusive}
\end{figure}

For assessing the energetics we note that, as stated in Sec.~\ref{subsec:heat}, no heat is dissipated from the system into the environment during a dispersive monitoring verifying the micro-reversibility relation in Eq.~\eqref{eq:ldb-diffusive}. This is in accordance with the fact that the interaction with the cavity preserves the qubit energy, and hence for every trajectory:
\begin{eqnarray}
\Delta E_\Lambda (\gamma_{[0,\tau]}) =  W_\Lambda (\gamma_{[0,\tau]}),
\end{eqnarray}
which {consequently} will only depend on the initial and final eigenstates of the trajectory $\gamma_{[0,\tau]}$.
The two different contributions to the work are in this case:
\begin{align}
W_\Lambda^{\mathrm{driving}}(\gamma_{[0,\tau]}) =& \tr[\Pi_n^0 ~V(0)] - \tr[\rho_\gamma(\tau) V(\tau)] \nonumber \\ & + i\omega_q \Omega_R \int_0^ \tau dt \langle \sigma_y \rangle_t \sin(\omega_q t),
\end{align}
where again the switch-on and switch-off interaction energy costs are identically zero. On the other hand, the measurement work reads in this case, from Eq.~\eqref{eq:meas-work-diffusive}:
\begin{eqnarray}
W_\Lambda^{\mathrm{meas}}(t) &=  \tr[H(\lambda_\tau) \Pi_m^{\tau}] - \tr[H(\lambda_\tau) \rho_\gamma(\tau)] \nonumber \\
&+ 2 \sqrt{k} \int_0^\tau dw(t) \left[ \langle H \sigma_z \rangle_t - E_\gamma(t) \langle \sigma_z\rangle_t \right],
\end{eqnarray}
with {again} $E_\gamma(t) = \langle H(\lambda_t) \rangle$ including both the qubit and driving Hamiltonians. In Fig.~\ref{fig:work-diffusive}a we compare the above driving and measurement contributions to the work for a single trajectory together with the total work $W_\Lambda(\gamma_{[0,\tau]})$. We observe that the total work (being equal to the energy changes in the qubit) shows again Rabi oscillations due to the driving, which however are not damped or intersected by abrupt changes anymore. Since the trajectory starts in the excited state, the total work is negative and oscillates between 0 (whenever the state of the system is again excited) and $-\omega_q$ (when the system state reaches the ground state) and hence $\omega_q$ is extracted. Nevertheless, the period of such oscillations becomes stochastic due to noise with a variance that increases over time, leading to the dephasing behavior when averaging over trajectories. The driving work is perfectly smooth in this situation and shows oscillations that reproduce the total work frequency combined with a slower modulation. Instead the measurement work captures all the energy fluctuations due to white noise in the evolution, that are associated to the disturbance in the state of the system due to the measurement.

Since the stochastic heat is zero for every trajectory, the entropy production in this case only accounts for the entropy changes in the system due to the (unital) measurement process, $S_\mathrm{tot}(\gamma_{[0,\tau]}) = \Delta S(\gamma_{[0,\tau]}) = - \log p_{n_\tau}^\tau + \log p_{n_0}^0$. In this case the entropy production admits a trivial split into adiabatic and non-adiabatic contributions. {Because of the unitality property of the map, the nonequilibrium potential operator $\Phi_\lambda = - \log \pi_\lambda$ verifies} $[\Phi_\lambda , L_z] = 0$ and hence {all} nonequilibrium potential changes {are zero}, $\Delta \phi_k (\lambda) = 0$ for all $k$ in Eq.~\eqref{eq:noneq-condition}. As a consequence the non-adiabatic entropy production in Eq.~\eqref{eq:detailed-na} just becomes the total entropy production, and the adiabatic entropy production in Eq.~\eqref{eq:detailed-ad} vanishes. Importantly, the latter does not imply that there might not be extra entropy production due to processes not related to the heat exchanged with the system, as it is the case e.g. in Ref.\cite{Popovic2021}. The split between uncertainty and martingale entropy productions is also simpler in this case and becomes $\Delta S_\mathrm{unc}(\gamma_{[0,\tau]}) = -\log(p_{n_\tau}^\tau) - S_\psi(\tau)$ and $\Delta S_\mathrm{mar}(\gamma_{[0,\tau]}) = \log(p_{n_0}^0) + S_\psi(\tau)$.

In Fig.~\ref{fig:work-diffusive}b we plot the total entropy production (blue curve) {for} a single trajectory and compare it with the martingale (``classicalized'') version (orange curve). As can be seen there, quantum fluctuations are dominant in this diffusive scenario since the changes in the entropy of the system due to the final virtual measurement are in general greater than the changes accumulated during the evolution. Blank spaces corresponds to intervals of time where the system state is very close to an eigenstate of $\rho(\tau)$ at that time. As time becomes comparable to the fully dephasing time, $1/\kappa$, the entropy change converges to a definite value which is independent of the final outcome, $\Delta S(\gamma_{[0,\tau]}) \rightarrow \log2 + \log p_{n_0}^0$ for $n_0 = \{ 0, 1\}$. Since the sample trajectory in the figure starts in the excited state ($n_0 = 1$), we have $\Delta S(\gamma_{[0,\tau]}) \rightarrow -0.62$ corresponding to $p_{0}^0 = 0.269$ the excited state probability at the initial time. On the other hand, the martingale entropy production shows small fluctuations due to the white noise contribution, similarly to the energy variation and the total work. It always remains {within} the envelope generated by the quantum fluctuations in $S_\mathrm{tot}(\gamma_{[0,\tau]})$ and coincides with it in the periods where the conditional state of the system is close to a eigenstate of the density matrix. 

\begin{figure}[tb]
    \centering
    \includegraphics[width=0.9 \linewidth]{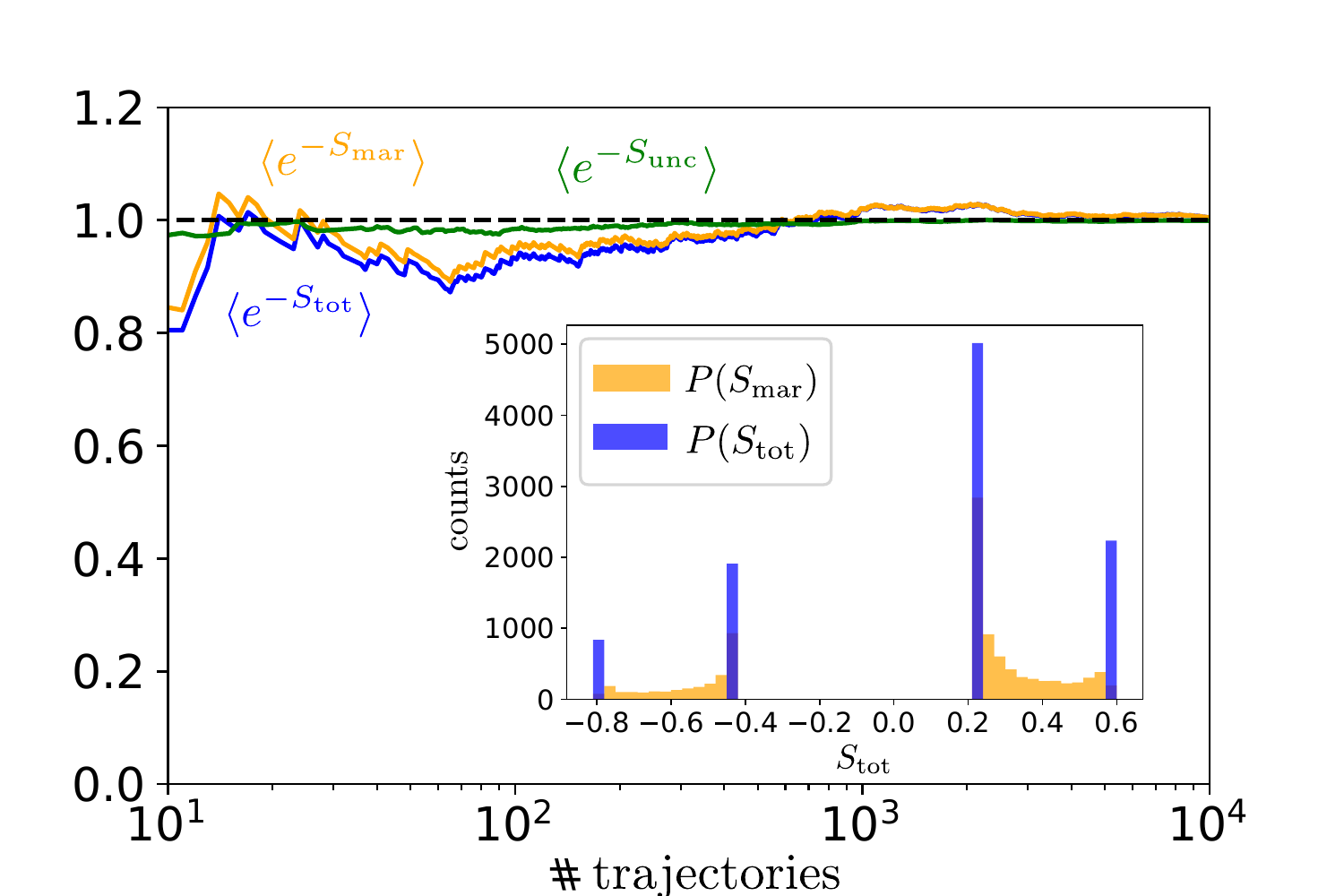}
    \caption{Integral fluctuation theorems for total $\langle e^{-S_\mathrm{tot}} \rangle_\gamma$ (blue), martingale $\langle e^{-S_\mathrm{mar}} \rangle_\gamma$ (orange) and uncertainty $\langle e^{-S_\mathrm{unc}} \rangle_\gamma$ (green) entropy production, as a function of the number of trajectories employed in the simulations for $\omega \tau = 1000$. Inset: Histograms of the total and martingale entropy productions for $N=10^4$ trajectories and $\omega \tau = 1000$, the estimated probability densities are obtained by dividing the number of counts over the total number of trajectories. 
    Other parameters of the simulation: $\kappa = 0.001 \omega$, $\Omega_R = 0.01 \omega_q$.
    }
    \label{fig:ft-diffusive}
\end{figure}

The stochastic evolution of the total and martingale entropy productions can be best appreciated in the inset of Fig.~\ref{fig:ft-diffusive}, where the probability distributions for $S_\mathrm{tot}(\gamma_{[0,\tau]})$ and $S_\mathrm{mar}(\gamma_{[0,\tau]})$ are plotted for a fixed final time, $\tau =1000 \omega_q^{-1} = \kappa^{-1}$. In particular we observe the 4 points in $P(S_\mathrm{tot})$, two on the left side corresponding to the cases where the system is initially in the excited state (and hence $S_\mathrm{tot}$ takes on negative values as in Fig.~\ref{fig:work-diffusive}b) and two positive corresponding to the case in which initially the qubit starts in the ground state. Moreover, as can be appreciated from $P(S_\mathrm{mar})$, the martingale entropy production can take continuous values between the maximum and minimum of $S_\mathrm{tot}$ for each initial state (but it cannot cross from positive to negative values and viceversa). The bimodal, highly non-Gaussian shape of the entropy production makes that the averages $\langle S_\mathrm{tot} \rangle_\gamma \geq 0$ and $\langle S_\mathrm{mar}\geq 0 \rangle_\gamma$ (or even they variances) are poorly informative of the actual dynamics of the system, and the effect of the measurement noise on the system. In any case we check in Fig.~\ref{fig:ft-diffusive} that the fluctuation theorems for both the total and martingale entropy productions are again verified as the functionals $\langle e^{-S_\mathrm{tot}} \rangle_\gamma$, $\langle e^{-S_\mathrm{mar}} \rangle_\gamma$, and $\langle e^{-S_\mathrm{unc}} \rangle_\gamma$ all tend to $1$ when increasing the number of trajectories in the simulation.

\section{Discussion and Outlook}

In this review we have discussed the application of the quantum trajectories framework for describing open quantum systems that are continuously monitored, to asses their thermodynamics. We conclude by stressing the key elements we used to accomplish a coherent description of all thermodynamic quantities at the stochastic level, namely, the introduction of a general TPM scheme on which quantum trajectories are embedded, the identification of entropy flow (and hence heat) from the micro-reversibility principle for quantum trajectories, Eq.~\eqref{eq:ef}, a suitable identification of the stochastic work induced by measurement back-action, Eqs.~\eqref{eq:meas-work} and \eqref{eq:meas-work-diffusive}, and the identification of entropy production from the likelihood of probabilities in forward and backward processes. For the latter point, it was important to define both processes in clear operational terms, that is, explicitly stating the initial states, driving protocol implemented and monitoring scheme implemented. The benefits of following such a recipe consist in recovering a coherent framework for the description of fluctuations of the main thermodynamic quantities, where central nonequilibrium relations associated to the second law, such as the fluctuation theorems, hold (as expected from an all-inclusive analyses~\cite{Campisi2011,Deffner2011,Manzano2018}).

From the experimental perspective, {despite many works have already explored the thermodynamics of average quantities in open systems,} still very few aspects of stochastic quantum thermodynamics have been tested in the laboratory. Some remarkable exceptions are for example the implementation of a Maxwell demon following diffusive trajectories in a circuit QED setup~\cite{Naghiloo2018}. The authors showed how the information acquired by monitoring a superconducting qubit system could be used to adequately implement a discrete feedback pulse for work extraction. We considered this setup in Sec.~\ref{sec:examples}B (without feedback), where other thermodynamic quantities not considered in Ref.~\cite{Naghiloo2018}, like the driving/measurement work split or the martingale entropy production could be also accessed, and their fluctuation relations tested. Another promising QED platform~\cite{Pekola2015} where most of the predictions of the framework for the case quantum jumps could be tested, consists of the implementation of extremely precise calorimetry on resistors acting as (finite-size) thermal reservoirs through fluorescence measurements~\cite{Karimi2020,Karimi2020b}.

Two other experiments tested the energetics of quantum monitored systems at an effective zero temperature~\cite{Naghiloo2020,Rossi2020}, a situation that involves many subtleties from the thermodynamic perspective and over which there is hence not complete consensus within the community. 
In Ref.~\cite{Naghiloo2020} a driven superconducting qubit is radiatively coupled to a transmision line whose electromagnetic field is subjected to a homodyne measurement by a Josephson parametric amplifier. Tracking the stochastic state of the system, the authors construct the total energy exchanged with the environment along single trajectories, and compare their results with a TPM scheme using energy measurements at different instants of time, and with a feedback loop that effectively isolates the system. Moreover, the authors compare tentative definitions of work and heat with the standard average expressions from the master equation, {obtaining good agreement}. Unfortunately, the experimental results are not sufficient for confirming or discarding a particular decomposition into heat and work along trajectories. In Ref.~\cite{Rossi2020} an optomechanical system was subjected to homodyne detection of the cavity field, leading to the monitoring of the nanomechanical oscillator Gaussian state at effective zero temperature. The focus was the assessment of the Wigner entropy production, and the authors were also able to identify the informational gain term due to the measurement. {Finally, a closely related situation has been considered in Refs.~\cite{Fabbri2020,Fabbri2021}, where fluctuation theorems has been tested in some particular situations using a nitrogen-vacancy (NV) center in diamond and an engineered dissipation channel. The system is subjected to repeated projective measurements (rather than following a continuous monitoring) while the authors could address driven-dissipative cases where only measurements of the system energy changes are needed to construct either heat or work~\cite{Fabbri2021}.}

We notice that situations with strictly zero temperature remain out of the formalism introduced here, since the condition for the Lindblad operators to include their adjoint pairs (or to be self-adjoint) would not be verified. However, we stress that such situations correspond to an idealization, since following the third-law of thermodynamics, attaining zero temperature would need infinite time~\cite{Levy2012,Levy2012b}, infinite dimensions~\cite{Silva2016} or infinite resources~\cite{Masanes2017,Taranto2021}. This situation can be solved {within the formalism} by allowing a small but non-zero temperature, accounting for the fact that adjoint processes (e.g. absorption of energy quanta from the environment) are improbable but not completely impossible.

In order to improve the applicability of the framework to most common experimental situations in different platforms, it would be desirable to systematically include the effects of non-efficient detectors in the thermodynamic framework. Moreover, the present approach for both quantum jumps and diffusive trajectories, might be extended to cases where not all Lindblad operators are included in the monitoring. In such cases, the expressions for the entropy flows above derived, and hence of heat and work, would need to be modified to take into account the flow of entropy not detected from the measured currents or jumps. Such an extension would be interesting in view of possible applications of the framework for e.g. entropy production estimation under hidden currents, in analogy to stochastic thermodynamics~\cite{Seifert2019}. A detailed analysis of such an extension is left for future work.

It is also worth mentioning that the results obtained within the quantum trajectory approach followed here are equivalent, in the case of quantum jumps, to multi-time correlation function approaches~\cite{Chetrite2012} and to the full counting statistics (FCS) method~\cite{Esposito2009,Kindermann2003}, although the difficulty to obtain and interpret the main thermodynamic quantities might be different. A comparison of those frameworks with the quantum jump method has been provided e.g. in Refs.~\cite{Liu2014b,Suomela2014,Liu2016b} for particular cases, {as well as with alternative methods~\cite{Solinas2015,Carrega2016}}. {In the FCS method, counting fields modeling the interaction between system and detectors in the reservoirs are introduced, which leads to a generalized master equation for a modified density operator that depends on these fields~\cite{Esposito2009,Kindermann2003}, and from which the moments of the heat currents can be obtained (see also Ref.~\cite{Bruderer2014}). This approach have been extended to the case of periodically driven systems combining it with Floquet theory~\cite{Gasparinetti2014,Kohler2016} and non-equilibrium Green function approaches~\cite{Kosov2020}, and it has been used to asses work, heat or efficiency fluctuations in thermoelectric systems~\cite{Galperin2015,Friedman2018}, quantum thermal machines~\cite{Agarwalla2017,Segal2018,Restrepo2018,Bouton2021}, to study Landauer's principle~\cite{Guarnieri2017} or to explore the consequences of TUR breakdown~\cite{Ptaszynski2018,Kalaee2021}}.
Connections of the quantum trajectory approach with collisional models used to describe the dynamics of open quantum systems, have been discussed in the recent review~\cite{Ciccarello2021}. {Moreover, recent developments used a collisional approach to asses information dynamics and their connection with thermodynamics in the continuously monitored scenario~\cite{Landi2022,Belenchia2022}.}

Finally, many works in recent years discussed the advantages and inconveniences of using TPM schemes, including both fundamental and practical considerations. On the fundamental side, one of the main critiques to the TPM scheme using projective energy measurements is that it is not suitable for considering initial states of the system bearing coherences in the energy basis~\cite{Perarnau2017} (but coherences developed during the system evolution are captured~\cite{Jarzynski2015}). Although often overlooked in the literature, this issue is avoided by allowing more general observables in the initial and final projective measurements of the TPM. When such observables coincide with the density operator of the unconditioned system the TPM does not cause indeed any disturbance on the system state{, as discussed before}. This approach can be complemented {by} introducing so-called augmented trajectories (or Bayesian networks), that include extra virtual measurements (e.g. over subsystems) without introducing their backaction~\cite{Park2020,Santos2019}, and have shown useful to obtain extra fluctuation relations~\cite{Micadei2020}. On the more practical side, although the TPM have been directly implemented in several  experiments~\cite{Kim2015,Xiong2018,Campisi2018,Zhang2018,Wu2019}, projective measurements are often difficult to control and may destroy the quantum system over which they are performed. Alternative  schemes reproducing their results have been hence proposed which reproduce the TPM statistics by using indirect measurements and interferometry techniques~\cite{Dorner2013,Mazzola2013,Goold2014,Roncaglia2014,Chiara2015,Talkner2016,Rubino2021}. Some of these alternative scheme have been successfully tested in the laboratory to measure work distributions and test fluctuation theorems in closed systems~\cite{Batalhao2014,Cersiola2017,Solfanelli2021}. We expect that such techniques may be easily extended to the case of more general observables than the energy, as considered here.

\begin{acknowledgments}
{
We acknowledge the Spanish State Research Agency QUARESC Project
(PID2019-109094GB-C21 /AEI/10.13039/501100011033), the Severo Ochoa and Maria de Maeztu Program for Centers and Units of Excellence in R \& D (MDM-2017-0711), and CAIB
 QUAREC Project (PRD2018/47). 
 G.M. acknowledges funding from Spanish MICINN through the 'Juan de la Cierva' program (IJC2019-039592-I).
 }
\end{acknowledgments}

\section*{AUTHOR DECLARATIONS}
\subsection*{Conflict of Interest}
The authors have no conflicts to disclose.

\appendix

\section{Derivation of work contributions with multiple conserved quantities.}
\label{app1}
In order to reach the decomposition of work (power) in Eq.~\eqref{eq:work-split} with the contributions from driving work in Eq.~\eqref{eq:drive-work}, chemical work in Eq.~\eqref{eq:work-chemical} and measurement work in Eq.~\eqref{eq:meas-work}, we start by subtracting the heat flux(es), that is work is identified as $\dot{W}_\Lambda(t) \equiv \dot{E}_\Lambda(t) - \sum_r \dot{Q}_\Lambda^{(r)}$, which from Eq.~\eqref{eq:currents} yields: 
\begin{align} \label{eq:power-app}
&\dot{W}_\Lambda(t) = \dot{W}_\Lambda^\mathrm{drive}(t) + \tr[H(\lambda_t) \dot{\rho}_\gamma(t)] - \sum_r \dot{Q}_\Lambda^{(r)}  = \dot{W}_\Lambda^\mathrm{drive}(t) \nonumber \\
&- \sum_k \tr[H \mathds{M}_k(\rho_\gamma)] + \sum_k \frac{dN_k}{dt} \tr[H \mathds{J}_k(\rho_\gamma)] - \sum_r \dot{Q}_\Lambda^{(r)},
\end{align}
where in the second line we expanded the middle term using the stochastic master equation as in Eq.~\eqref{eq:intermediate}. 

We will now manipulate the middle term in the second line of \eqref{eq:power-app} associated to the quantum jumps, $\tr[H \mathds{J}_k(\rho_\gamma)] = \tr[L_k^\dagger H L_k \rho_\gamma]/\langle L_k^\dagger L_k \rangle - E_\gamma$ by using the commutation relations between the Hamiltonian $H$ and the Lindblad operators $L_k$. For this purpose, and for remaining as general as possible, we split the Hamiltonian as $H = H_S + V$, where $H_S$ is the part of the system energy which verifies energy conservation within system and reservoir [and hence enters as a charge in Eq.~\eqref{eq:ladder-ops}] and $V$ is a weak interaction. Hence we split the term $\tr[L_k^\dagger H L_k \rho_\gamma] = \tr[L_k^\dagger H_S L_k \rho_\gamma] + \tr[L_k^\dagger V L_k \rho_\gamma]$. Crucially, we now rewrite Eq.~\eqref{eq:ladder-ops} for the entropy flow with multiple conserved quantities as:
\begin{eqnarray} \label{eq:commutation-app}
[H_S, L_k] = - T_r \Delta S_k^{(r)} L_k + \sum_{i>1} \nu_i^{(r)}[X_i, L_k], 
\end{eqnarray}
where we simply assumed $X_1 = H_S$ (energy conservation) with $\nu_1^{(r)} = -1$ and rearranged terms. Notice that here above $k$ represents a channel that belongs to reservoir $r$ with which energy and the other charges $\{X_i\}$ for $i>1$ are exchanged. This equation allow us to manipulate the expression $\tr[L_k^\dagger H_S L_k \rho_\gamma]$ for such a channel by using the following two equations directly coming from \eqref{eq:commutation-app}:
\begin{align}
H_S L_k &= L_k H_S - T_r \Delta S_k^{(r)} L_k + \sum_{i>1} \nu_i^{(r)}[X_i, L_k] , \\
L_k^\dagger H_S &= H_S L_k^\dagger - T_r \Delta S_k^{(r)} L_k + \sum_{i>1} \nu_i^{(r)}[X_i, L_k],
\end{align}
which introduced into $\tr[L_k^\dagger H_S L_k \rho_\gamma]$ give us the two following equivalent relations:
\begin{align} \label{eq:rel1-app}
\tr[L_k^\dagger H_S L_k \rho_\gamma] =& \tr[L_k^\dagger L_k H_S \rho_\gamma] - \langle L_k^\dagger L_k \rangle T_r \Delta s_k^{(r)} \nonumber \\ &+ \sum_{i>1} \nu_i^{(r)}\tr[L_k^\dagger [X_i, L_k]\rho_\gamma], \\ \label{eq:rel2-app}
\tr[L_k^\dagger H_S L_k \rho_\gamma] =& \tr[H_S L_k^\dagger L_k \rho_\gamma] - \langle L_k^\dagger L_k \rangle T_r \Delta s_k^{(r)} \nonumber \\ & + \sum_{i>1} \nu_i^{(r)}\tr[[L_k^\dagger, X_i] L_k]\rho_\gamma].
\end{align}
Combining the r.h.s. of Eqs.~\eqref{eq:rel1-app} and \eqref{eq:rel2-app} with equal $1/2$ weights then leads to the expression that we wanted:
\begin{align} \label{eq:rel-app}
 \tr[L_k^\dagger H_S L_k \rho_\gamma] =& \frac{1}{2} \tr[\{L_k^\dagger L_k, H_S\} \rho_\gamma] - \langle L_k^\dagger L_k \rangle T_r \Delta s_k^{(r)} \nonumber \\ & + \sum_{i>1} \nu_i^{(r)}\tr[\mathds{D}_k(\rho_\gamma)],
\end{align}
where we recall that channel $k$ belongs to reservoir $r$. Introducing Eq.~\eqref{eq:rel-app} into the quantum jump term proportional to $\tr[H \mathds{J}_k(\rho_\gamma)]$ in Eq.~\eqref{eq:power-app} and performing the sum over $k$ for the different reservoirs, we immediately obtain:
\begin{align}\label{eq:jumps-app}
&\sum_k \frac{dN_k}{dt} \tr[H \mathds{J}_k(\rho_\gamma)] = \sum_k \frac{dN_k}{dt} \Big( \frac{\tr[L_k V L_k \rho_\gamma]}{\langle L_k^\dagger L_k \rangle} \nonumber \\ 
&+ \frac{1}{2} \frac{\tr[\{L_k^\dagger L_k, H_S\} \rho_\gamma]}{\langle L_k^\dagger L_k \rangle} - E_\gamma \Big) + \sum_r \dot{Q}_\Lambda^{(r)} + \dot{W}_\Lambda^\mathrm{chem},
\end{align}
where we identified the heat currents $\dot{Q}_\Lambda^{(r)}= - T_r \sum_k (dN_k/dt) \Delta s_k^{(r)}$ coming from the second term in Eq.~\eqref{eq:rel-app}, and the chemical power $\dot{W}_\Lambda^\mathrm{chem}$ in Eq.~\eqref{eq:work-chemical}, coming from the last contribution in Eq.~\eqref{eq:rel-app}.

Introducing Eq.~\eqref{eq:jumps-app} into the power split of Eq.~\eqref{eq:power-app} the heat current contributions cancel and we obtain:
\begin{align}
&\dot{W}_\Lambda(t) = \dot{W}_\Lambda^\mathrm{drive}(t) + \dot{W}_\Lambda^\mathrm{chem} - \sum_k \tr[H \mathds{M}_k(\rho_\gamma)] \\ &+ \sum_k \frac{dN_k}{dt} \Big( \frac{\tr[L_k V L_k \rho_\gamma]}{\langle L_k^\dagger L_k \rangle} + \frac{1}{2} \frac{\tr[\{L_k^\dagger L_k, H_S\} \rho_\gamma]}{\langle L_k^\dagger L_k \rangle} - E_\gamma \Big). \nonumber
\end{align}
Finally, Eq.~\eqref{eq:work-split} is recovered by identifying the remaining terms in the above equation with the measurement work:
\begin{align} \label{eq:meas-work-app}
\dot{W}_\Lambda^\mathrm{meas} \equiv& - \sum_k \tr[H \mathds{M}_k(\rho_\gamma)] + \frac{dN_k}{dt} \Big( \frac{\tr[L_k^\dagger V L_k \rho_\gamma]}{\langle L_k^\dagger L_k \rangle} \nonumber \\ &+ \frac{1}{2} \frac{\tr[\{L_k^\dagger L_k, H_S\} \rho_\gamma]}{\langle L_k^\dagger L_k \rangle} - E_\gamma \Big),
\end{align}
where we recall that $\mathds{M}_k(\rho_\gamma) = {\frac{1}{2}}\{L^\dagger_k L_k, \rho_\gamma\} - \tr(L_k^\dagger L_k \rho_\gamma) \rho_\gamma$. We notice that in the case $H_S = H$ and $V=0$, we recover Eq.~\eqref{eq:meas-work}. Moreover, assuming $V$ is of the order of the coupling between system and reservoirs, terms like $\tr[V \mathds{M}_k(\rho_\gamma)] \sim O(||V||^3)$ can be neglected, and we can replace $H$ by $H_S$ in the first term of the r.h.s. in the above equation. However, the second (jump) term  proportional to the weak interaction $V$ during the jumps remains (since it is normalized by $\langle L_k^\dagger L_k \rangle$). By noticing that $E_\gamma = \tr[H \rho_\gamma] = \tr[H_S \rho_\gamma] + \tr[V \rho_\gamma]$, we may interpret the extra term in Eq.~\eqref{eq:meas-work-app}:
\begin{eqnarray} \label{eq:meas-work-interaction}
\sum_k \frac{dN_k}{dt} \Big( \frac{\tr[L_k^\dagger V L_k \rho_\gamma]}{\langle L_k^\dagger L_k \rangle} - \tr[V \rho_\gamma] \Big) \equiv \dot{W}_\Lambda^{\mathrm{int}},
\end{eqnarray}
as the work needed to maintain the interaction $V$ connected to the system (or to instantaneously switching it off and on again) when local jumps occur. This contribution should only be neglected at the average level where $\langle dN_k \rangle_\gamma = dt \langle L_k^\dagger L_k \rangle$ compensates for the normalization and hence $\tr[L_k^\dagger V L_k \rho] \sim \langle L_k^\dagger L_k \rangle \tr[V \rho_\gamma] \sim O(||V||^3)$, and hence $\langle \dot{W}_\Lambda^{\mathrm{int}} \rangle_\gamma = 0$. Therefore although no net work is needed to maintain the interaction $V$ while the system (locally) exchanges energy with the reservoirs, this nevertheless induces extra power fluctuations as quantified by Eq.~\eqref{eq:meas-work-interaction}. 

\section{Derivation of the adiabatic and non-adiabatic decomposition for quantum stochastic entropy production.}
\label{app}

In this appendix we provide more details on the decomposition into adiabatic and non-adiabatic contributions of the entropy production for monitored systems (Sec.~\ref{sec:ep}). As mentioned there, we introduce dual and dual-reversed processes that are similar to the forward and backward process, but employ a modified set of Lindlbad operators. We denote these sets by $\{ L_k^{+} \}_{k = 1}^K$ and $\{\tilde{L}_k^{+} \}_{k = 1}^K$ respectively, which verify~\cite{Horowitz2013,Manzano2018b}:
\begin{align} \label{eq:dual-jumps}
     L_k^{+}(\lambda_t) ~~&=~ e^{\frac{1}{2}(\Delta s_k + \Delta \phi_k)} ~L_k^\dagger(\lambda_t), \\ \label{eq:dualr-jumps}
     \tilde{L}_k^{+}(\lambda_{\tau - t}) &=~ e^{\Delta \phi_k/2} ~\Theta L_k^\dagger(\lambda_t) \Theta^\dagger.
\end{align}
On the other hand the Hamiltonian part of the evolution is the same as in forward and backward processes,  $H(\lambda)$ and $\tilde{H}(\lambda)$, respectively. That is, the control protocol $\Lambda$ is implemented in the dual process  and the time-reversed control protocol $\tilde{\Lambda}$ is implemented in the dual-reverse process. Moreover, the above equations also guarantee that the instantaneous steady states in the dual and dual-reverse dynamics coincide with those in the forward and backward dynamics, respectively.

In this context it is useful to define the trajectory operator that generates the environmental record $\gamma_{(0,\tau)}$ in the dual process. We denote it $\pazocal{T}_\Lambda^{+}(\gamma_{(0,\tau)}) =  \mathcal{U}^{+}(\tau, t_J) L_{k_J}^{+} ... L_{k_1}^{+} \mathcal{U}^{+}(t_1, t_0)$, where the no jump intervals are governed by the same drift evolution as in the forward process, $\mathcal{U}^{+}(t_j, t_i) = \mathcal{U}(t_j, t_i)$ and the dual jumps are given in Eq.~\eqref{eq:dual-jumps} in terms of the original ones. Analogously, we denote the operator associated to the trajectory $\tilde{\gamma}_{[0,\tau]}$ in the dual-reverse process as $\pazocal{T}_{\tilde{\Lambda}}^{+}(\tilde{\gamma}_{(0,\tau)}) =  \tilde{\mathcal{U}}^{+}(\tau, t_J) \tilde{L}_{k_J}^{+} ... \tilde{L}_{k_1}^{+} \tilde{\mathcal{U}}^{+}(t_1, 0)$ with $\tilde{\mathcal{U}}^{+}(t_j, t_i) = \Theta \mathcal{U}^\dagger(\tau - t_j, \tau- t_i) \Theta^\dagger$ and $\tilde{L}_k^+$ in Eq.~\eqref{eq:dualr-jumps}. These operators are related to the probabilities of observing the trajectories $\gamma_{[0,\tau]}$ and $\tilde{\gamma}_{[0,\tau]}$ in the dual and dual-reversed process respectively:
\begin{align}
P_\Lambda^{+} &= p_{n_0}^0 \tr[\Pi_{n_\tau}^\tau \pazocal{T}_\Lambda^{+}(\gamma_{(0,\tau)}) \Pi_{n_0}^\tau \pazocal{T}_\Lambda^{+ ~\dagger}(\gamma_{(0,\tau)})], \\
P_{\tilde{\Lambda}}^{+} &= p_{n_\tau}^\tau \tr[\tilde{\Pi}_{n_0}^0 \tilde{\pazocal{T}}_\Lambda^{+}(\tilde{\gamma}_{(0,\tau)}) \tilde{\Pi}_{n_0}^\tau \tilde{\pazocal{T}}_\Lambda^{+ ~\dagger}(\tilde{\gamma}_{(0,\tau)})],
\end{align}
to be compared with Eqs.~\eqref{eq:prob} and \eqref{eq:prob-b}. Using the relations in Eqs.~\eqref{eq:dual-jumps}-\eqref{eq:dualr-jumps} for the dual and dual-reversed jumps, together with the corresponding ones for the drift evolution operators, we obtain the following relations:
\begin{align} \label{eq:micro-dual}
 \pazocal{T}_\Lambda^{+ \dagger}(\gamma_{(0,\tau)}) &= \pazocal{T}_{\Lambda}({\gamma}_{(0,\tau)}) ~e^{\frac{1}{2}[\Delta \Phi_\Lambda({\gamma}_{(0,\tau)}) + \sigma_\Lambda({\gamma}_{(0,\tau)})]}, \\ \label{eq:micro-dualr}
 \Theta^\dagger  \pazocal{T}_{\tilde \Lambda}^{+ \dagger}(\tilde{{\gamma}}_{(0,\tau)}) \Theta &=   \pazocal{T}_{\Lambda}({\gamma}_{(0,\tau)}) ~e^{\Delta \Phi_\Lambda({\gamma}_{(0,\tau)})/2},
\end{align}
where we introduced the accumulated change in nonequilibrium potential during the whole trajectory:
\begin{eqnarray}
\Delta \Phi_\Lambda({\gamma}_{(0,\tau)}) = \sum_{j=1}^J \Delta \phi_{k_j}(\lambda_{t_j}). 
\end{eqnarray}
The relations \eqref{eq:micro-dual}-\eqref{eq:micro-dualr} are analogous to the micro-reversibility relation in Eq.~\eqref{eq:ef} but help us to relate the statistics of the dual and dual-revesed processes to the original forward process, which are the key to obtain the detailed fluctuation theorems in Eqs.~\eqref{eq:detailed-ad} and \eqref{eq:detailed-na}.

On a more technical side, we note that in Ref.~\cite{Fagnola2007} it was shown that conditions equivalent to Eq.~\eqref{eq:noneq-condition} on the Lindblad operators (called a privileged representation) are verified when the so-called $s$-dual generator $\tilde{\mathcal{L}}^+_\lambda(\rho) \equiv \pi_\lambda^{1-s} \mathcal{L}_\lambda^\ast(\pi^{s} \rho \pi^{1-s}) \pi_\lambda^{s}$ associated to $\mathcal{L}_\lambda$, for $s \in [0,1]$ commutes with the modular automorphism, $\mathcal{M}(\rho) = \pi_\lambda \rho \pi_\lambda^{-1}$ (see Theorem 8 and Proposition 19 in Ref.~\cite{Fagnola2007}). This constraint in turn ensures that the maps generated by $\tilde{\mathcal{L}}^+_\lambda(\rho)$ are unique and form a quantum Markov semigroup $\forall~ s \in [0,1]$, while the symmetric dual generator ($s=1/2$) always leads to a quantum Markov semigroup even when commutation with the modular automorphism is not ensured (see Theorem 36 and Example 41 in Ref.~\cite{Fagnola2007}). The condition in Eq.~\eqref{eq:noneq-condition} allowing the entropy production split and the derivation of separate fluctuation theorems for the two pieces are equivalent to the existence of a privileged representation for the symmetric dual generator in the case of jump trajectories~\cite{Horowitz2013}. However such conditions, as introduced in Refs.~\cite{Manzano2015,Manzano2018b}, ensure the split and the fluctuation theorems for general CPTP maps, not necessarily forming a quantum Markovian semigroup.

\section*{DATA AVAILABILITY}
Data sharing is not applicable to this article as no new data were
created or analyzed in this study.

\bibliography{biblio}

\end{document}